\documentclass[fleqn,usenatbib]{mnras}

\usepackage{newtxtext,newtxmath}
\usepackage[T1]{fontenc}

\DeclareRobustCommand{\VAN}[3]{#2}
\let\VANthebibliography\thebibliography
\def\thebibliography{\DeclareRobustCommand{\VAN}[3]{##3}\VANthebibliography}

\usepackage{graphicx}	% Including figure files
\usepackage{amsmath}	% Advanced maths commands
\title[J0140: an evolved CV turned ELM WD]{LAMOST J0140355+392651: An evolved cataclysmic variable donor transitioning to become an extremely low mass white dwarf}

\author[El-Badry et al.]{Kareem El-Badry,$^{1,3}$\thanks{E-mail: kelbadry@berkeley.edu}
Eliot Quataert,$^{2,1}$
Hans-Walter Rix,$^{3}$
Daniel R. Weisz,$^{1}$
Thomas Kupfer,$^{4}$  \newauthor 
Ken Shen,$^{1}$
Maosheng Xiang,$^{3}$
Yong Yang,$^{5}$
Xiaowei Liu$^{6}$
\\
$^{1}$Department of Astronomy and Theoretical Astrophysics Center, University of California Berkeley, Berkeley, CA 94720, USA\\
$^{2}$Department of Astrophysical Sciences, Princeton University, Princeton, NJ 08544, USA\\
$^{3}$Max-Planck Institute for Astronomy, K\"onigstuhl 17, D-69117 Heidelberg, Germany\\
$^{4}$Department of Physics \& Astronomy, Texas Tech University, P.O. Box 41051, Lubbock, TX 79409, USA\\
$^{5}$South-Western Institute for Astronomy, Yunnan University, Kunming 650500, P. R. China\\
$^{6}$Department of Astronomy, Peking University, Beijing
100871, P. R. China \\
}

\date{Submitted to MNRAS}

\pubyear{2021}

\begin{document}
\label{firstpage}
\pagerange{\pageref{firstpage}--\pageref{lastpage}}
\maketitle

\begin{abstract}
We present LAMOST J0140355+392651 (hereafter J0140), a close ($P_{\rm orb} = 3.81$ hours) binary containing a bloated, low-mass ($M \approx 0.15 M_{\odot}$) proto-white dwarf (WD) and a massive ($M \approx 0.95\,M_{\odot}$) WD companion. The system's optical light curve is dominated by large-amplitude ellipsoidal variability but also exhibits additional scatter, likely driven by pulsations. The proto-WD is cooler ($T_{\rm eff} = 6800\pm 100$ K) and more puffy ($\log\left[g/\left({\rm cm\,s^{-2}}\right)\right]=4.74\pm0.07$) than any known extremely low mass (ELM) WD, but hotter than any known cataclysmic variable (CV) donor. It either completely or very nearly fills its Roche lobe ($R/R_{{\rm Roche\,lobe}}=0.99\pm0.01$), suggesting ongoing or recently terminated mass transfer. No dwarf nova-like outbursts have been observed. The spectrum is dominated by the proto-WD but shows tentative hints of H$\alpha$ emission, perhaps due to accretion onto the massive WD. The properties of the system are well-matched by MESA binary evolution models of CVs with donors that underwent significant nuclear evolution before the onset of mass transfer. In these models, the bloated proto-WD is either still losing mass via stable Roche lobe overflow or was doing so until very recently. In either case, it is evolving toward higher temperatures at near-constant luminosity to become an ELM WD. If the system is detached, mass transfer likely ended when the donor became too hot for magnetic braking to remain efficient. Evolutionary models predict that the binary will shrink to $P_{\rm orb}\lesssim 10$ minutes within a few Gyr, when it will either merge or become an AM CVn binary. J0140 provides an observational link between the formation channels of CVs, ELM WDs, detached ultracompact WD binaries, and AM CVn systems. 
\end{abstract}


%%%%%%%%%%%%%%%%%%%%%%%%%%%%%%%%%%%%%%%%%%%%%%%%%%

%%%%%%%%%%%%%%%%% BODY OF PAPER %%%%%%%%%%%%%%%%%%

\section{Introduction}

Extremely low mass white dwarfs (ELM WDs, with masses  $M\lesssim 0.25 M_{\odot}$) are degenerate and semi-degenerate helium stars that never ignited core helium burning. The Universe is too young to produce ELM WDs by single-star evolution. It is thus thought that all ELM WDs are products of binary evolution: they are the stripped cores of stars that were initially more massive ($M_{\rm init}\gtrsim 1\,M_{\odot}$) but lost most of their envelope to a companion \citep[e.g.][]{Iben_1986}. This envelope stripping can occur through common envelope evolution or through stable mass transfer \citep[e.g.][]{Li2019}. In the stable mass transfer channel, short-period ELM WDs are descendants of cataclysmic variables (CVs) with evolved donor stars \citep[e.g.][]{Kalomeni_2016}. When mass transfer ceases, the donor shrinks inside its Roche lobe and appears as a detached ELM WD. When and why mass transfer ceases is not fully understood, but is presumably linked to the end of efficient angular momentum loss via magnetic braking \citep[e.g.][]{Sun_2018}.

Consistent with predictions of the mass transfer scenario, essentially all known ELM WDs are in binaries. Their companions are normal WDs and neutron stars \citep{Brown2020}. ELM WDs are rare: their local space density is about $5\times 10^{-5}$ that of ordinary WDs \citep{Brown2016}. Most of the $\sim$100 spectroscopically confirmed ELM WDs were discovered in the last decade by the dedicated ELM survey \citep{Brown2010}, which performed a comprehensive magnitude-limited search beginning with SDSS color cuts. Known ELM WDs are relatively hot ($T_{\rm eff}\gtrsim 8,000$\,K) and compact ($\log g \gtrsim 5.5$). This is largely a consequence of the color cuts adopted by the ELM survey: cooler and more bloated proto-ELM WDs are expected to exist but are not easily distinguishable from main-sequence A and F stars based on their colors alone \citep[e.g.][]{Pelisoli2018, Pelisoli2018b, Pelisoli2019b}. 

The mass-transferring progenitors of short-period ELM WDs differ from ordinary CVs in that the donor stars underwent significant nuclear evolution before the onset of mass transfer. In most short-period ($P_{\rm orb} \lesssim 6$ hours) CVs, the mass-losing stars fall on a tight ``donor sequence'' of mass, radius, spectral type, and luminosity as a function of period \citep{Patterson_1984, Beuermann_1998, Smith_1998, Knigge_2006, Knigge_2011}. The regularity of the CV donor population is a consequence of the fact the donor stars fill their Roche lobes (so that their mean density is a nearly deterministic function of orbital period) and are only mildly out of thermal equilibrium. They thus have radii and effective temperatures similar to main-sequence stars of the same mass. 

CVs donors that were significantly evolved at the onset of Roche lobe overflow (RLOF) are predicted to deviate from the donor sequence  \citep[e.g.][]{Tutukov_1985, Podsiadlowski_2003, vandersluys2005, Kalomeni_2016} because they have already formed a helium core when mass transfer commences. By the time the orbital period reaches $P_{\rm orb}\lesssim 6$ hours, such donors will consist of a low-mass ($M\lesssim 0.2 M_{\odot}$) helium core with a thin hydrogen-burning envelope. The structure of such a donor is quite different from that of a main-sequence star of the same mean density. The most obvious observable difference is that the donor stars in CVs formed from evolved donors are expected to be hotter and smaller than normal CV donors with the same period. Several CVs with potentially evolved donors have been identified  \citep[e.g.][]{Augusteijn_1996, Thorstensen_2002, Thorstensen_2002b, Littlefair_2006, Thorstensen_2013, Kato_2014, Rebassa_2014, Ashley2020}. Most of these systems have K-type donor stars, sometimes with abundances bearing imprints of CNO burning, at periods where normal CVs have M-type donors. If the donors are sufficiently evolved, they can become detached ELM WDs. 

CVs with evolved donors are of interest for several reasons. Shortly after initial RLOF, their mass transfer rate is often high enough to allow for steady hydrogen burning on the surface of the accreting WD. This can form a bright supersoft X-ray source \citep[e.g.][]{vandenHeuvel1992} and, if a high mass transfer rate is sustained long enough, trigger a type Ia supernova \citep[e.g.][]{Han2004}. If magnetic braking remains efficient, evolved CVs can become mass-transferring ultra short-period ($5 \lesssim P_{\rm orb}/{\rm minute} \lesssim 30$) ``AM-CVn'' binaries containing a helium WD donor \citep[e.g.][]{Podsiadlowski_2003}. If magnetic braking becomes inefficient (e.g., because the donor loses its convective envelope), evolved CVs become ELM WDs, which can eventually become ultracompact via gravitational wave radiation \citep[e.g.][]{Kilic2014, Burdge2020}.

This paper presents a newly discovered close binary containing a normal white dwarf and a low-mass companion that is nearly or completely Roche lobe filling. The lower-mass component appears to be in a rarely observed transitional state between evolved CV donors and ELM WDs. It is hotter than any known CV donors, but cooler than any known ELM WDs. 
The optical spectrum is dominated by the low-mass component, which has an early F spectral type, and contains only hints of emission lines. No dwarf novae outbursts are detected; if they occur, their recurrence timescale would be unusually long. Because there is no strong evidence of ongoing mass transfer, the system either is detached or has an unusually low mass transfer rate.  These unusual properties warrant detailed study, as the binary serves as a unique observational link between  populations of CVs, ELM WDs, and ultracompact binaries.

The remainder of this paper is organized as follows. Section~\ref{sec:data} summarizes the observed data, including archival spectra, light curves from several time-domain photometric surveys, astrometry from {\it Gaia}, the UV-to-near infrared spectral energy distribution (SED), and follow-up radial velocity observations.  We constrain the physical parameters of the binary using these data in Section~\ref{sec:constraints} and derive an upper limit on the outburst frequency in Section~\ref{sec:no_outburst}.  In Section~\ref{sec:mesa}, we present binary evolution models for J0140 and discuss the future evolution of the system. We summarize our results and discuss future observations in Section~\ref{sec:discussion}. Supporting data is provided in the appendices. 

\section{Data}
\label{sec:data}
\subsection{LAMOST discovery spectra}
\label{sec:lamost}

J0140 first came to our attention because LAMOST observations  \citep{Cui_2012}  of the object suggested large epoch-to-epoch radial velocity (RV) variability. The object was visited by LAMOST three times, yielding two usable spectra (see Table~\ref{tab:lamost}). Between these two visits, the reported RV varied by $\approx 400$\,$\rm km\,s^{-1}$.

Most LAMOST ``visits'' consist of 2-4 consecutive ``subexposures'', with individual exposure times of 10-45 minutes. In the default LAMOST data reduction pipeline, these subexposures are coadded to improve SNR, without any velocity shift. The stellar parameters and RVs reported in most LAMOST catalogs are inferred from the coadded spectra. For close binaries, RVs may change significantly between subexposures, and the coadded spectra may have artificially broadened lines due to orbital smearing. We therefore investigated the RVs for J0140 across subexposures of each visit (Appendix ~\ref{sec:spec_details} and Figure~\ref{fig:lamost_rvs}), finding RV variations of up to 600\,$\rm km\,s^{-1}$ within a single visit. This prompted us to obtain additional follow-up spectra, as described in Section~\ref{sec:kast}.

\subsection{ZTF light curve}
\label{sec:ZTF} 

J0140 was observed regularly by the public survey of the Zwicky Transient Facility  \citep[ZTF;][]{Bellm_2019, Graham_2019, Masci_2019}. We queried the public DR5 light curves for all clean photometry (\texttt{catflags} = 0). This yielded 524 $g$-band exposures, 576 $r$-band exposures, and 42 $i$-band exposures. The exposure time is 30 seconds in all bands. The ZTF data span 964 days between 2018 and 2021, sometimes including several observations on a single night.

\begin{figure*}
    \centering
    \includegraphics[width=\textwidth]{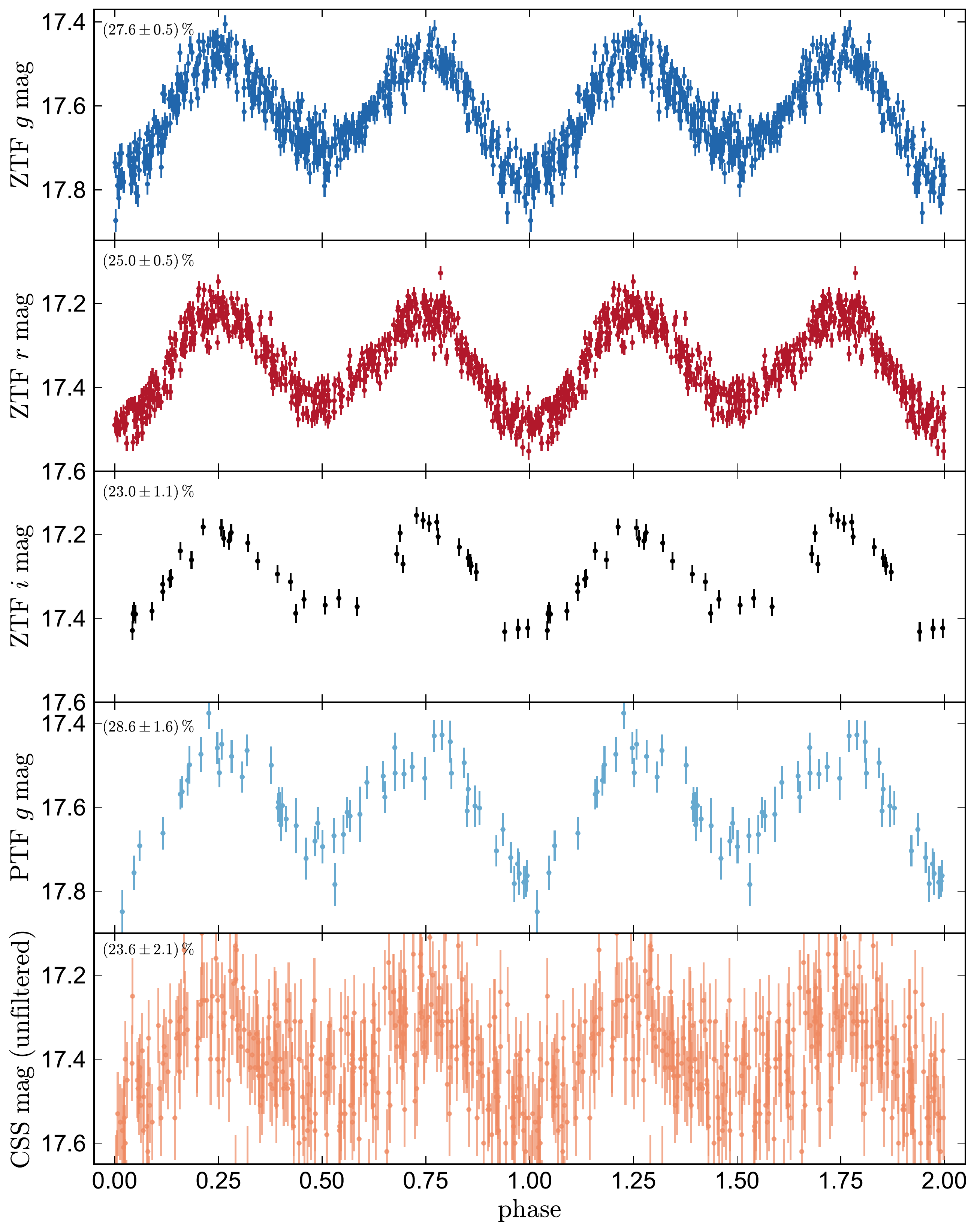}
    \caption{ZTF, PTF, and CSS light curves of J0140, phased to a period of 3.81 hours. Each datapoint appears twice. The $y$-axis spans 0.55 mag in all panels. Variability is dominated by ellipsoidal modulation. Constraints on the peak-to-peak flux variability amplitude are listed in the upper left of each panel.
    %The variability amplitude increases slightly toward bluer bands.
    }
    \label{fig:ztf_lcs}
\end{figure*}

To estimate the orbital period of J0140, we calculated $g-$ and $r-$ band periodograms from the ZTF light curves. Both were dominated by a peak at 1.904 hours. In binaries whose light curves are dominated by ellipsoidal variation, the dominant light curve period is typically half the orbital period, so we folded the light curves on twice this period (Figure~\ref{fig:ztf_lcs}). This revealed a light curve shape characteristic of ellipsoidal modulation: quasi-sinusoidal variability with alternation in the depth of adjacent minima. This alternation stems from the fact that a tidally distorted star is fainter when its cooler end faces the observer (phase 0 or 1.0) than when its hotter end does (phase 0.5).

\subsection{Other photometry}
\label{sec:other_lc}

\begin{figure*}
    \centering
    \includegraphics[width=\textwidth]{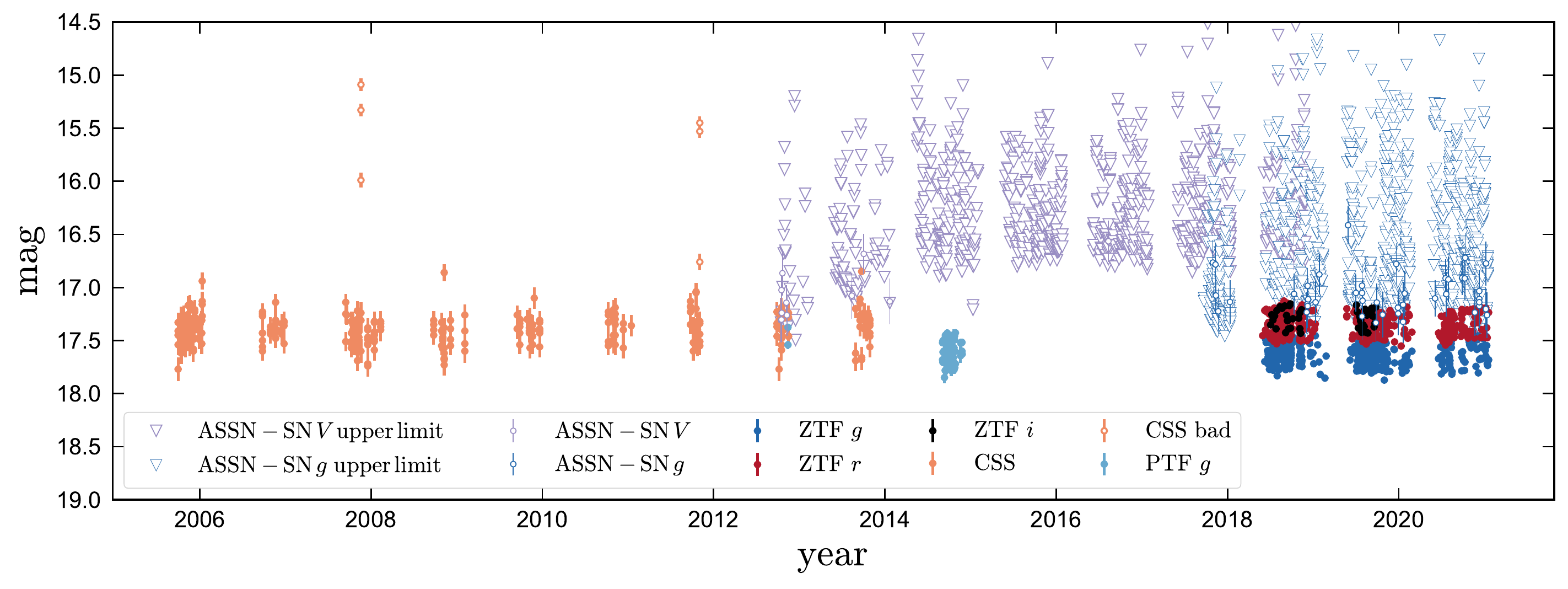}
    \caption{Light curve of J0140, showing data from several time-domain surveys. Points with errorbars are detections; inverted triangles are  upper limits. CSS magnitudes are unfiltered but are scaled to roughly approximate the $V$ band. The two apparent outbursts in the CSS light curve (in 2007 and 2011) are spurious; no reliable dwarf-novae like outbursts were detected between 2005 and 2021.  }
    \label{fig:full_lc}
\end{figure*}

J0140 was also observed by several other time-domain photometric surveys. 
Figure~\ref{fig:full_lc} shows all the photometric data we analyze, including both detections and upper limits between 2005 and 2021. The highest-quality photometry is from ZTF, but ZTF photometry only covers a small fraction of the total time baseline. Phased light curves from three surveys are shown in Figure~\ref{fig:ztf_lcs}.

\subsubsection{CSS}
The Catalina Sky Survey \citep[CSS;][]{Drake_2009} observed J0140 regularly between 2005 and 2013. CSS magnitudes are unfiltered but are transformed to $V$ using an empirically calibrated, color-dependent relation. We queried all clean CSS photometry for J0140 with photometric uncertainty $\sigma_V < 0.12$ mag. This yielded 330 photometric points with a typical uncertainty of 0.09 mag. 

The phased CSS light curve exhibits similar ellipsoidal modulation to the ZTF light curves (Figure~\ref{fig:ztf_lcs}). The full light curve (Figure~\ref{fig:full_lc}) also contains two episodes, in 2007 and 2011, in which the object appears to brighten by 2-3 mag. These episodes at first appear similar to outbursts found in dwarf novae due to the disk instability \citep[e.g.][]{Osaki1996}. However, closer inspection revealed that within each episode, the unusually bright light curve points (with magnitude 15-16) were flanked on either side by observations with normal brightness (magnitude 17-18), with temporal separations of only a few minutes. Since dwarf novae outbursts typically last several days, this aroused suspicion that the apparent brightening might instead be due to problems with the photometry. We retrieved the raw CSS images  for the 2011 episode from the CRTS web portal\footnote{\href{https://crts.iucaa.in/CRTS/}{https://crts.iucaa.in/CRTS/}} and found that the brightening is indeed spurious. Images for the 2007 brightening episode are not available via the CRTS portal, but given their erratic time-evolution, we suspect this episode is also due to data problems and is not a dwarf nova outburst.  

\subsubsection{PTF}
\label{sec:ptf}
J0140 was also observed by the Palomar Transient Factor \citep[PTF;][]{Law_2009}. The light curve contains 86 good datapoints with a typical uncertainty of 0.04 mag, including 71 measurements in the $g$ band and and 15 in the $R$ band. The exposure time is 60 seconds. The phased PTF light curve exhibits similar ellipsoidal modulation to the ZTF and CSS light curves (Figure~\ref{fig:ztf_lcs}). We limit our analysis to the $g$ band data.

\subsubsection{ASAS-SN}
J0140 is regularly observed by the The All-Sky Automated Survey for Supernovae \citep[ASAS-SN;][]{Kochanek_2017}, but the object is generally too faint to be detected. We retrieved $V-$ and $g-$band light curves from the sky-patrol web portal\footnote{https://asas-sn.osu.edu/} on MJD 59232. The light curve contains 992 $V$ band datapoints and 1030 $g$ band datapoints, but the majority are upper limits (Figure~\ref{fig:full_lc}). There are 12 $V$ band detections and 43 $g$ band detections, both with a median magnitude of 17.1. The brightest nominal detection has a $g$ band magnitude of $16.41\pm 0.18$.

Most ASAS-SN exposures are collected in sets of three consecutive exposures, each with an exposure time of 90 seconds. In all the ASAS-SN detections, there was a detection in only one of the three exposures in the set, usually with a magnitude similar to the upper limits reported for the other two exposures. It is thus unlikely that any of the anomalously bright ASAS-SN detections are dwarf novae outbursts. 

\subsection{Light curve analysis}
\label{sec:lc_analysis}
\subsubsection{Variability amplitude}
\label{sec:var_amp}
To constrain the photometric variability amplitude, we fit the photometry from each survey shown in Figure~\ref{fig:ztf_lcs} with a 3-term Fourier model; i.e., modeling the light curve in normalized flux space as a sum of sines and cosines with periods $P$, $P/2$, and $P/3$, and the phases held fixed. In fitting these amplitudes, we also model a photometric scatter term that accounts for intrinsic scatter and/or underestimated photometric uncertainties. The resulting constraints on the peak-to-peak variability amplitude are shown in the upper left of each panel of Figure~\ref{fig:ztf_lcs}. Differences between the filters are subtle, but there is a slight trend of larger variability amplitude at shorter wavelengths. This could be due either to wavelength-dependence in the intrinsic variability amplitude of the star (i.e. due to wavelength-dependent gravity darkening), or to a wavelength-dependent flux ratio between the star and a non-variable component.

\subsubsection{Photometric scatter}
\label{sec:scatter_phot}

\begin{figure}
    \centering
    \includegraphics[width=\columnwidth]{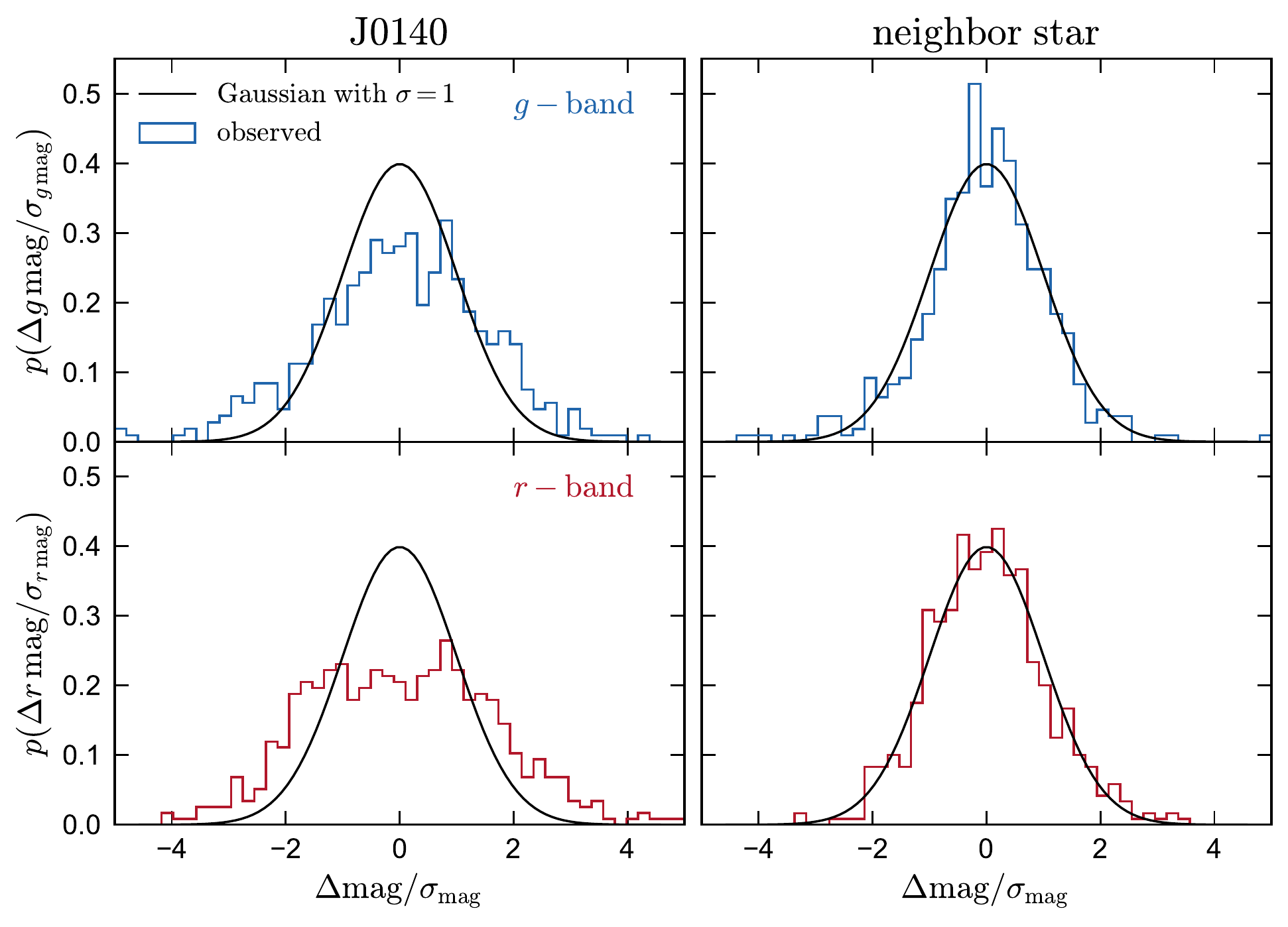}
    \caption{Difference between observed ZTF magnitudes and the mean magnitude at the corresponding orbital phase, normalized by the reported magnitude uncertainties. We compare distributions for J0140 (left) to those for a nearby comparison star with similar magnitude that is not significantly variable (right). For J0140, the distribution is broader than a Gaussian with $\sigma=1$. This suggests that either (a) there is additional intrinsic scatter in the light curve, or (b) the magnitude uncertainties are underestimated. The well-behaved Gaussian distribution with $\sigma\approx 1$ for the neighbor star suggests that additional intrinsic scatter is the more likely explanation. This scatter is most likely due to pulsations of the proto-WD.  }
    \label{fig:deltamag}
\end{figure}

The ZTF $g$- and $r$-band light curves in Figure~\ref{fig:ztf_lcs} appear to show somewhat more scatter than expected from the reported observational uncertainties. This is illustrated in Figure~\ref{fig:deltamag}, which shows the uncertainty-normalized residuals between the observed magnitudes and the mean magnitude at their phase, which is calculated by binning the observed light curves in phase. If the magnitude were only a function of orbital phase, with no intrinsic scatter or dependence on other parameters, and the reported magnitude uncertainties were reliable, these residuals would be expected to follow a Gaussian distribution with $\sigma =1$. The fact that the observed distribution is wider than this implies that either the magnitude uncertainties are underestimated, or there is additional variability. 

The distribution of photometric residuals for a non-variable neighbor star with similar magnitude to J0140 is indeed well described by a Gaussian with $\sigma=1$, suggesting that the reported uncertainties are reliable. We do not find strong evidence of any additional periodicity in the residuals; it appears to be stochastic. The origin of this additional variability is unknown. One possibility is that the proto-WD pulsates, as several ELM WDs are observed to \citep[e.g.][]{Hermes2012, Maxted2014, Gianninas2016}. Indeed, the temperature and surface gravity of proto-WD (Section~\ref{sec:kast}) place it within the pre-ELM WD instability strip \citep[e.g.][]{Corsico2016}. It is also possible that the variability is due to ``flickering'' of an accretion disk around the massive WD, which is observed in almost all CVs \citep[e.g.][]{Bruch_1992} and is thought to be a result of turbulence in the disk.

To quantify the amplitude of the additional scatter, we fit the $g-$ and $r-$band light curves with a Fourier series (see Section~\ref{sec:ephemeris}). Along with the Fourier amplitudes, we fit a scatter term, $\sigma_{\rm scatter}$, which is added in quadrature to the observational uncertainties in the likelihood function. We find $\sigma_{\rm scatter}=2.4\pm 0.1 \%$ in the $r$ band and $2.4\pm 0.1 \%$ in the $g$ band. Higher-cadence photometry of the binary is required to better characterize this variability and verify whether it is due to pulsations.

\subsubsection{Orbital ephemeris}
\label{sec:ephemeris}
We determined the orbital ephemeris by simultaneously fitting the normalized CSS, PTF, and ZTF light curves with a harmonic series. The long temporal baseline covers more than 35,000 orbits and thus allows for a high-precision determination of the orbital period (Table~\ref{tab:system}). The quoted conjunction time ``$T_0$'' is the time corresponding to phase 0, when the companion (almost certainly a WD) passes in front of the proto-WD. When we analyze the radial velocities in Section~\ref{sec:kast}, we find good agreement with the photometric ephemeris. This unambiguously establishes the photometric variability as tracing the orbital period of the binary. 

We do not find evidence of orbital decay: fitting the full light curve and allowing for a time-dependent phase shift, we find $\dot{P}_{{\rm orb}}=-4_{-3}^{+5}\times10^{-11}\,{\rm s\,s^{-1}}$, consistent with 0. This non-detection is consistent with the theoretically expected value from gravitational waves for the parameters derived in Section~\ref{sec:constraints}, which is $\dot{P}_{{\rm orb,GW}}=-7.7_{-6.4}^{+3.4}\times10^{-14}\,{\rm s\,s^{-1}}$. If magnetic braking remains efficient (see Section~\ref{sec:mesa_mdoels}), the orbital decay due to it is expected to be significantly faster than this, $\dot{P}_{{\rm orb,MB}}\approx -2.5\times10^{-12}\,{\rm s\,s^{-1}}$, still consistent with the non-detection.

\subsection{KAST Spectra}
\label{sec:kast}
We observed J0140 using the KAST double spectrograph on the 3\,m Shane telescope at Lick observatory in July and August 2020. Most of our observations used the 600/7500 grating on the red side and the 600/4310 grism on the blue side, with the D55 dichroic and a 2 arcsec slit. This setup results in a wavelength range of 3300–8400 \AA\,\,with typical resolution (FWHM) of 4 \AA\,\,on the blue side and 5 \AA\,\,on the red side. Some additional spectra were obtained with a separate setup (``B''). For these observations, we used the 1200/5000 grating on the red side, the 830/3460 grism on the blue side, the D46 dichroic, and a 1 arcsec slit. This setup results in a wavelength range of 3300–7200 \AA,\,\, with a gap in coverage at 4600–5400 \AA. The typical resolution for this setup is 3 \AA\,\,on the blue side and 2 \AA\,\,on the red side. %For both setups, we took 600 second exposures.

To minimize effects of flexure on the wavelength solution, a new set of arcs was taken on-sky while tracking the target every $\sim$30 minutes. The spectra were reduced using \texttt{pypeit} \citep{Prochaska_2020}, which performs bias and flat field correction, cosmic ray removal, wavelength calibration, flexure correction using sky lines, sky subtraction, extraction of 1d spectra, and heliocentric RV corrections. 

To balance the need to minimize orbital smearing with the need to reach high enough SNR to measure RVs at each epoch, we used 600 second exposures. This resulted in a typical per-pixel SNR of 10-15 on the red side and 5-10 on the blue side. The spectra are summarized in Table~\ref{tab:specs}. The quoted SNR is the median per pixel on the red side; the typical SNR on the blue side is a factor of $\sim$2 lower.
Variations in SNR between visits are due primarily to variations in sky brightness and airmass. 

\subsubsection{Radial velocities}
\label{sec:RVs}
Comparison of the single-epoch spectra with a grid of ATLAS/SYNTHE spectral models \citep{Kurucz_1970, Kurucz_1979, Kurucz_1993} suggested an effective temperature of $T_{\rm eff}\approx 6500\,\rm K$ and a surface gravity $\log( g/{\rm cm\,s^{-1}}) \approx 4.5$. We used a model spectrum with these parameters and $\rm [Fe/H]=0$ as a template to estimate the RVs of each single-epoch spectrum via cross-correlation.\footnote{Once we derived more accurate atmospheric parameters for the donor from the co-added spectrum (Section~\ref{sec:spec}), we updated the template and re-measured RVs. } RVs were measured from the red-side spectra, which have a more stable wavelength solution due to a larger number of sky emission lines. The most prominent spectral feature on the red side is the H$\alpha$ absorption line.

\begin{figure}
    \centering
    \includegraphics[width=\columnwidth]{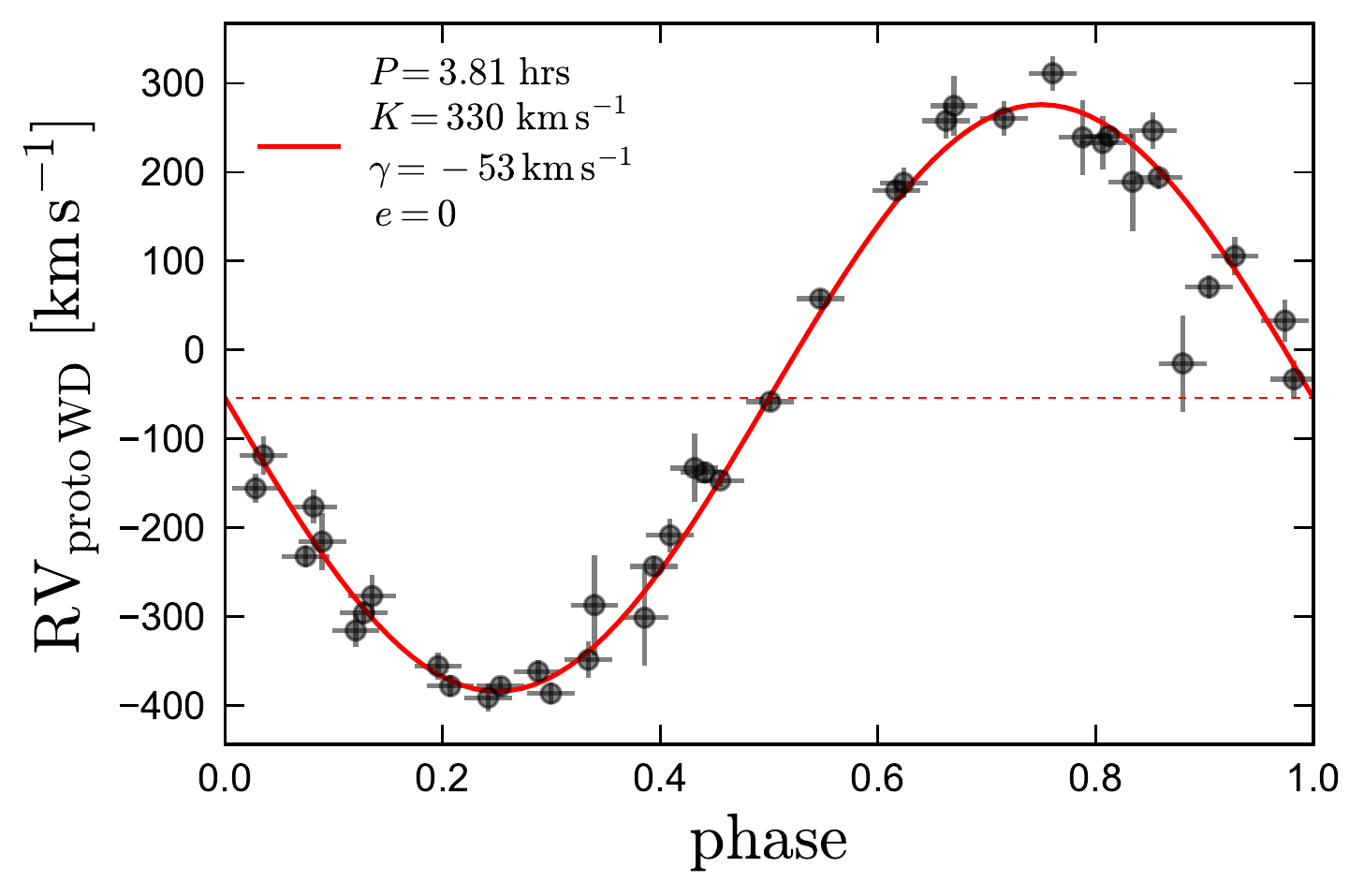}
    \caption{Radial velocity of the proto-WD, phased to a period of 3.81 hours. RVs are measured from Kast spectra (Table~\ref{tab:specs}), primarily from the H$\alpha$ absorption line. Horizontal error bars show the extent of each 10 minute exposure; vertical error bars are $1\sigma$ measurement uncertainties. Red line shows the best-fit Keplerian orbit (Table~\ref{tab:system}).  }
    \label{fig:rvs}
\end{figure}

\begin{table}
\centering
\caption{Basic observables of J0140. Uncertainties are  1$\sigma$ (middle 68\%). Additional constraints are in Table~\ref{tab:constraints}.}
\begin{tabular}{lll}
\hline\hline

\multicolumn{3}{l}{\bf{System parameters}}  \\ Right ascension (J2000) & $\alpha$\,[deg] & 14.065468 \\
Declination (J2000) & $\delta$\,[deg] & 39.447740 \\
Apparent magnitude & $G$ & 17.35 \\
Color & $G_{BP}-G_{\rm RP}$ & 0.63 \\
{\it Gaia} eDR3 parallax & $\varpi$\,[mas] & $0.65\pm 0.10$  \\
Zeropoint-corrected parallax  & $\varpi$\,[mas] & $0.68\pm 0.10$  \\ 
Reddening (SFD; to infinity) & $E(B-V)$ & 0.056  \\ 
Reddening (Bayestar2019) & $E(g-r)$ & $0.04\pm 0.02$  \\ 
\hline
\multicolumn{3}{l}{\bf{Parameters of the proto-WD}}  \\ 
Effective temperature (spectrum) & $T_{\rm eff}$\,[K] & $6800 \pm 100$ \\
Surface gravity   & $\log(g/(\rm cm\,s^{-2}))$  & $4.7\pm0.3$  \\
Metallicity    & $ [\rm Fe/H]$  & $-0.1\pm0.1$  \\
Projected rotation velocity & $v\sin i$\,[km\,s$^{-1}$] &  $105\pm 20$ \\
Angular diameter & $\Theta\,[\mu \rm as]$ & $2.05 \pm 0.05$ \\
Effective temperature (SED) & $T_{\rm eff}$\,[K] & $6820 \pm 110$ \\

\hline
\multicolumn{3}{l}{\bf{Parameters of the WD}}  \\ 
Angular Diameter & $\Theta\,[\mu \rm as]$ & $0.055 \pm 0.013$ \\
Effective temperature & $T_{\rm eff}$\,[K] & $19700 \pm 1500$ \\

\hline 
\multicolumn{3}{l}{\bf{Orbital parameters}}  \\ 
Period   & $P_{\rm orb}$\,[days] & 0.1586339(2)  \\ 
Conjunction time  & $T_0$\,[HMJD UTC] & 59060.490(1)  \\ 
RV semi-amplitude & $K\,[\rm km\,s^{-1}]$ & $330\pm 5$ \\
Center-of-mass RV & $\gamma\,[\rm km\,s^{-1}]$ & $-53\pm 4$ $\pm 30$ (sys) \\ 
eccentricity & $e$ & $<0.017$ \\
RV scatter & $s\,[\rm km\,s^{-1}]$ & $18\pm 5$\\
Mass function & $f_m\,[M_{\odot}]$ & $0.591\pm 0.027$ \\

\hline
\end{tabular}
\begin{flushleft}

\label{tab:system}
\end{flushleft}
\end{table}

\begin{figure}
    \centering
    \includegraphics[width=\columnwidth]{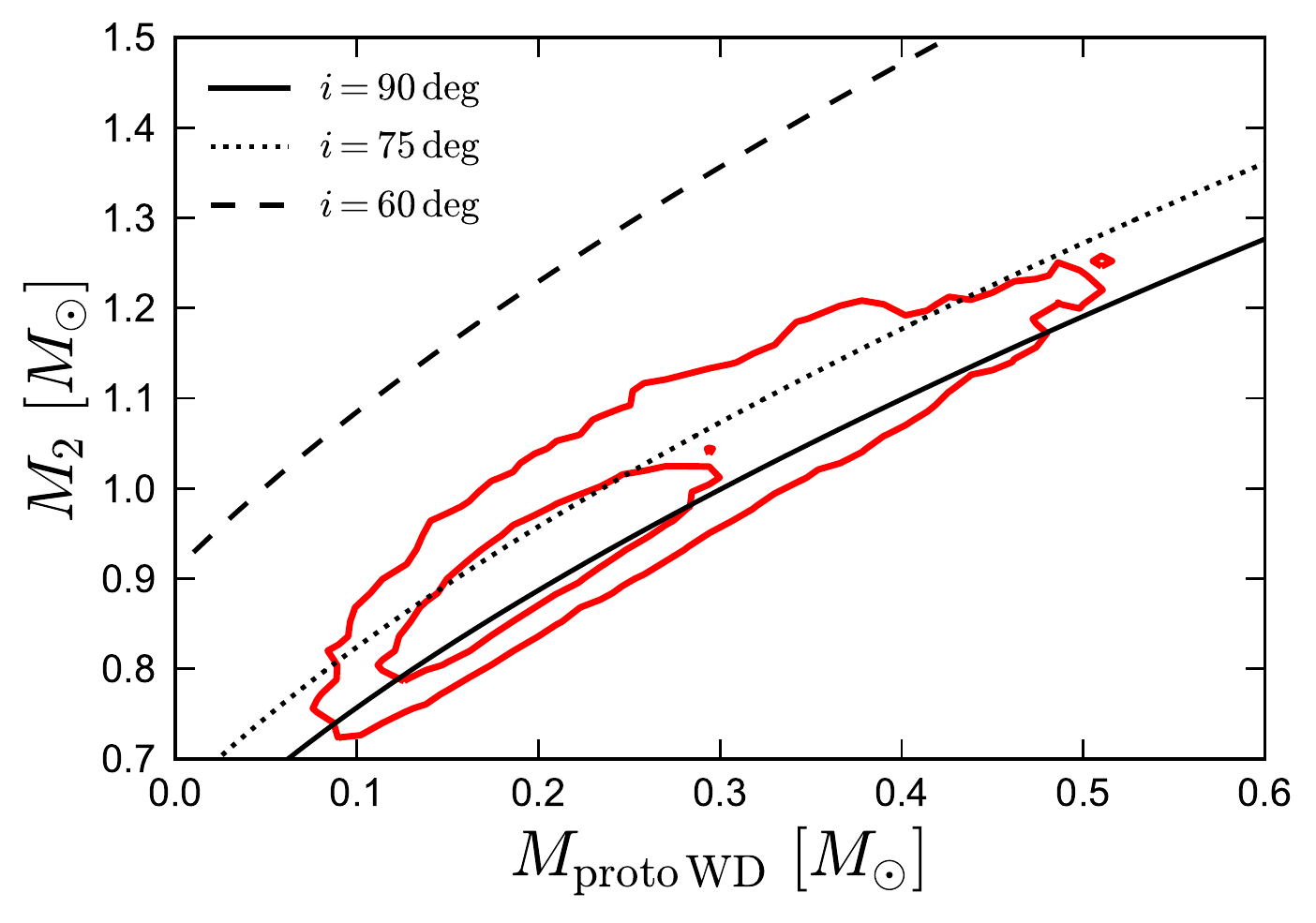}
    \caption{Dynamically implied mass of the companion as a function of the proto WD's mass and orbital inclination, given the observed mass function, $f_m \approx 0.59 M_{\odot}$. Red contours show 1 and 2 sigma constraints from joint fitting of the light curve, spectra, orbit, SED, and parallax (Section~\ref{sec:constraints}). Given the observed UV excess (Figure~\ref{fig:SED}), the $\approx 0.9 M_{\odot}$  companion is very likely a white dwarf.}
    \label{fig:masses}
\end{figure}

We fit an orbital solution to the RVs (which presumably trace the proto-WD) while holding the period fixed to the value measured from the light curve. 
Along with the 6 standard Keplerian orbital parameters, we fit for an intrinsic RV scatter or ``jitter'' term (see \citealt{Elbadry_2018}, their Equation 9), which can represent either intrinsic RV variability not captured in the model or underestimated uncertainties. We first search for the best-fit orbital solution using simulated annealing, and then sample from the posterior in the vicinity of this solution using a Markov chain Monte Carlo method, as described in \citet{Elbadry_2018}. Our best-fit orbital parameters are listed in Table~\ref{tab:system}. The eccentricity is consistent with 0. The mass function, $f_{m}=PK_{\rm proto\,WD}^{3}(1-e^{2})^{3/2}/(2\pi G)$, which represents the minimum possible mass of the companion, i.e., the mass it would have if the proto-WD were a test particle and the system were viewed edge on, is consistent with a white dwarf companion. This is illustrated in Figure~\ref{fig:masses}, which shows the dynamically implied masses of the companion for a range of proto-WD masses and inclinations.

The best-fit jitter term is $\approx 18\,\rm km\,s^{-1}$. Nonzero jitter is evident in Figure~\ref{fig:rvs}, where the scatter in RV measurements is somewhat larger than the 1$\sigma$ RV uncertainties suggest it should be. This jitter is most likely due to instability in the wavelength solution; it could also originate in part from the ambiguity in the definition of RVs for an ellipsoidally distorted star \citep[e.g.][]{Masuda2021}, from irradiation effects \citep[e.g.][]{Warner_2003}, or from pulsations. 

\subsubsection{Proto-WD spectrum}
\label{sec:spec}
We reconstructed a higher-SNR spectrum of the proto-WD by coadding spectra from all epochs, shifted to the star's rest frame. This yielded a combined spectrum with maximum $\rm SNR \approx 60$ per pixel and mean spectral resolution $R\approx 1,200$ (FWHM). We fit for the atmospheric parameters, projected rotation velocity, and metallicity of the donor using standard full spectral fitting based on a forward-model of ab-initio spectra. Rotational broadening was accounted for using the kernel from \citet{Gray1992}. 

We implemented {\it The Payne} \citep{Rix_2016, Elbadry_2018, Ting_2019}, a framework for interpolating model spectra, to predict the pseudo-continuum normalized spectra of the proto-WD. We use a 2nd order polynomial spectral model with three labels,\footnote{We also explored the possibility of determining the proto-WD's detailed abundance pattern, since constraining the abundance of C, N, and O would provide a useful test of the evolutionary scenario proposed in Section~\ref{sec:mesa}. We found lines from these elements to be too blended for reliable abundance measurements. } $\vec{\ell}=\left(T_{{\rm eff}},\log g,\left[{\rm Fe/H}\right]\right)$, which we trained on the \texttt{BOSZ} grid of Kurucz model spectra  \citep{Bohlin_2017}. In the relevant part of parameter space, the grid spacing is $\Delta T_{\rm eff} = 250\,\rm K$, $\Delta \log g = 0.5\, \rm dex$, $\Delta [\rm Fe/H] = 0.25\,\rm dex$. 
%This does {\it not} set the precision of the derived labels, since the model can interpolate between grid points. 
For both the models and the observed spectrum, we define the pseudo-continuum as a running median with width 150\,\AA. Low-resolution spectra are subject to severe line blending, so this pseudo-continuum has little to do with the true continuum. The purpose of pseudo-continuum normalization is simply to bring the observed and model spectra to the same scale, minimizing uncertainty due to unknown detector efficiency and distance. This fitting yields an effective temperature for the proto-WD of $T_{\rm eff} \approx 6800\,\rm K$, surface gravity $\log\left(g/{\rm \left[cm\,s^{-2}\right]}\right) \approx 4.7$, and metallicity $[\rm Fe/H] \approx -0.1$ (see Table~\ref{tab:system}).

The formal fitting uncertainties on spectroscopic labels are very small: 25 K in $T_{\rm eff}$, 0.06 dex in $\log g$, and 0.03 dex in $[\rm Fe/H]$. Except at very low-SNR, formal uncertainties derived from full-spectrum fitting underestimate the true parameter uncertainties \citep[e.g.][]{Ness2015, Xiang2019}. This owes to errors in pseudo-continuum normalization, simplifications in the spectral model (e.g., fitting only 3 labels) and systematics in the model spectra on which the models are trained. We therefore inflate the uncertainties (Table~\ref{tab:system}) to realistic values for the resolutions and SNR of the spectrum  \citep[e.g.][]{Xiang2019}.

%It is natural to wonder whether weak emission lines from a disk could preferentially dilute the Balmer lines, confounding our estimate of the effective temperature and flux ratio. We consider this unlikely, since a Kurucz model provides a reasonably good fit over most of the optical spectrum, including the Balmer lines, Ca H \&K lines, and various metal lines. If there {\it were} weak and unrecognized emission in the Balmer lines, it would tend to bias the fit toward lower $T_{\rm eff}$, since the strength of the donor's Balmer absorption lines increases with increasing $T_{\rm eff}$ in this temperature range.  

The strongest features in the proto-WD spectrum are the Balmer and calcium H and K lines. Other metal lines are also present, but they are subject to significant blending at low resolution. The agreement between the observed and model spectra is satisfactory for most lines. Some low-frequency undulations are evident in the residual spectrum; these likely are due to imperfect normalization. The most significant feature in the residuals is in the H$\alpha$ absorption line, which is weaker in the observed spectrum than in the model. This may be the result of contamination from weak H$\alpha$ emission originating in accretion onto the more massive WD: as we show in Section~\ref{sec:constraints}, the proto-WD is extremely close to Roche lobe-filling, so ongoing mass transfer is not unlikely. The hints of emission are, however, much weaker than found in typical CVs \citep[e.g.][]{Warner_2003}, and there is no evidence that a disk contributes significantly to the optical SED (Section~\ref{sec:sed}). If there is ongoing mass transfer, the accretion rate is likely much lower than in typical CVs at similar periods. 

The observed spectrum also has significantly stronger sodium ``D'' absorption lines (at 5890 and 5896 \AA) than the model. We observe these lines to shift coherently with the orbit of the proto-WD and can thus rule out interstellar absorption as their source. Sodium enhancement is frequently observed in evolved CV donors \citep[e.g.][]{Thorstensen_2002} and is thought to originate from CNO-processing in the core of the secondary near the end of its main-sequence evolution. In this scenario, sodium-enhanced material only reaches the surface of the secondary after most of the envelope has been stripped off. A similar evolutionary history can likely explain the formation of J0140 (Section~\ref{sec:mesa}).

\subsubsection{Variability with orbital phase}
\label{sec:phase_variability}

We also investigated whether the spectra vary with orbital phase (Appendix~\ref{sec:phased_specs}). This reveals weak evidence of increased H$\alpha$ absorption at $\phi =0.5$, as might be expected if a H$\alpha$-emitting component were partially or fully eclipsed, or if the WD-facing side of the proto-WD were irradiated by the more massive WD. However, the apparent increase in the line's depth is only $\approx$10\%. Further spectroscopic follow-up with higher SNR is required to confirm whether it is real.

\subsection{Distance}
\label{sec:distance}
The {\it Gaia} eDR3 parallax \citep{Gaia2020} of J0140 (Gaia DR3 source id 373857386785825408) is relatively significant, $\varpi = 0.65 \pm 0.10$. Interpreted as an inverse distance, this places J0140 at a distance of $d\approx 1.54$\,kpc (1-sigma range of 1.33 to 1.84\,kpc). The parallax zeropoint predicted by the fitting function from \citet{Lindegren2020zpt} for J0140 is -0.03 mas. At $G> 17$, {\it Gaia} eDR3 parallax uncertainties for blue sources are not overestimated on average \citep{Elbadry2021}. We thus adopt $\varpi = (0.68 \pm 0.1)$ in modeling the SED. The geometric distance from \citet{BailerJones2020}, which is based on the zeropoint-corrected parallax and a Milky Way model prior, is $d=1.51_{-0.18}^{+0.25}\,{\rm kpc}$.

The {\it Gaia} eDR3 astrometric solution does not account for photocenter wobble induced by orbital motion, but this is expected to be negligible: the ratio of the projected semimajor axis to the parallax, $\theta/\varpi= a/(\rm 1 au)$, is only about 0.005. The renormalized unit weight error (RUWE) is 0.97, indicating an unproblematic astrometric solution.

\begin{figure*}
    \centering
    \includegraphics[width=\textwidth]{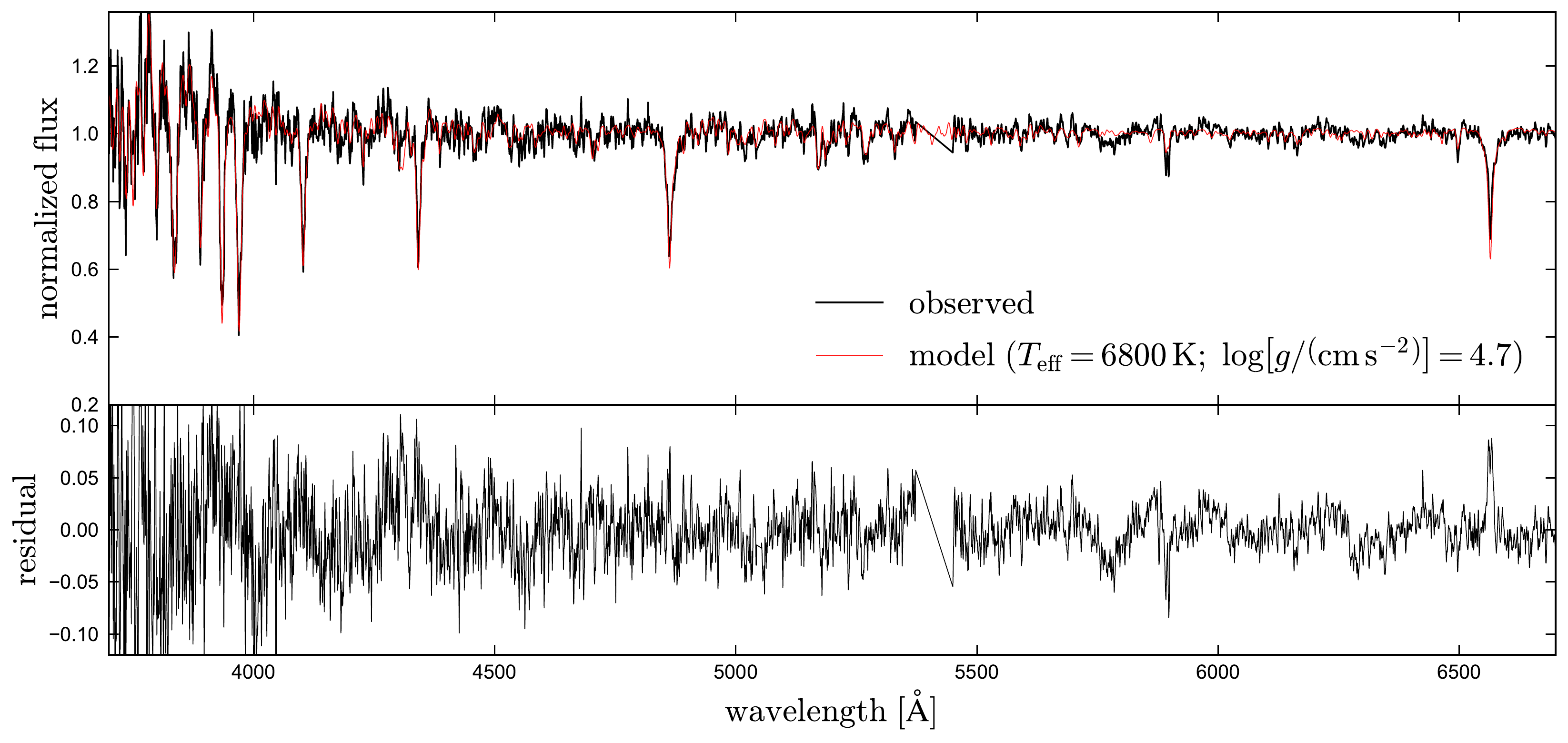}
    \caption{Rest-frame normalized spectrum of the proto-WD (black) and best-fit Kurucz model spectrum (red). The strongest absorption features are the Balmer lines and Ca H\,\&\,K lines.  Bottom panel shows residuals. A satisfactory fit is achieved with $T_{\rm eff} = 6800\,\rm K$, $\log = 4.7$, and $[\rm Fe/H] = -0.1$. Most of the structure in the residuals is broad and due to imperfect normalization. The most significant narrow feature in the residuals is in the H$\alpha$ line. This could arise from emission from a faint accretion disk around the WD, but may also be the result of systematics in the spectral model.  }
    \label{fig:spectra}
\end{figure*}

\subsection{Spectral energy distribution}
\label{sec:sed}
The spectral energy distribution (SED) of J0140 provides a constraint on the temperature and angular diameter of both components. Table~\ref{tab:sed} summarizes the available broadband fluxes, which are taken from the GALEX \citep{Morrissey_2007}, SDSS \citep{York_2000}, 2MASS \citep{Skrutskie_2006}, and WISE \citep{Wright_2010} surveys. Because J0140 exhibits ellipsoidal variability with amplitude of $\approx 0.3$ mag in the optical, we ``correct'' the reported magnitudes to better approximate their phase-averaged mean values. We use a Fourier model of the ZTF $r-$band light curve (Figure~\ref{fig:ztf_lcs}) to calculate a ``optical ellipsoidal correction'' (Table~\ref{tab:sed}) that converts the magnitude observed at a given phase to the mean magnitude in that band.

\begin{table*}
	\centering
	\caption{Spectral energy distribution. Magnitude lower limits are $2\sigma$. mjd values are mid-exposure. The WISE magnitudes represent a mean from 33 exposures spread roughly uniformly in phase. The ``corrected magnitude'' column corrects the observed magnitudes, which occur at different orbital phases, to represent the phase-averaged mean magnitude. This column also corrects for discrepancies between the SDSS and AB magnitude systems (Section~\ref{sec:sed}). }
	\label{tab:sed}
	\begin{tabular}{ccccccc} 
		\hline
		filter & $\lambda_{\rm eff}\,[\mu \rm m]$ & mag & mjd & phase & optical ellipsoidal correction [mag] & corrected\,\,mag \\
		\hline
		FUV & 0.154 & $21.90 \pm 0.13$ & 53682.4345 & $0.70 \pm 0.04$ & $0.11\pm 0.01$  & $21.90 \pm 0.13$ \\
		NUV & 0.227 & $20.55 \pm 0.04$ & 53682.4345 & $0.70 \pm 0.04$ & $0.11\pm 0.01$  & $20.63 \pm 0.08$ \\
		$u$ & 0.359 & $18.72 \pm 0.02$ & 54010.4614 & $0.52\pm 0.04$ & $-0.09\pm 0.03$  & $18.59 \pm 0.05$ \\
		$g$ & 0.464 & $17.64 \pm 0.02$ & 54010.4630 & $0.53 \pm 0.04$ & $-0.08\pm 0.03$ & $17.56 \pm 0.04$ \\
		$r$ & 0.612 & $17.40 \pm 0.01$ & 54010.4597 & $0.51 \pm 0.04$ & $-0.09\pm 0.02$ & $17.31 \pm 0.03$\\
		$i$ & 0.744 & $17.34 \pm 0.01$ & 54010.4605 & $0.52 \pm 0.04$ & $-0.09\pm 0.02$ & $17.27 \pm 0.03$ \\
		$z$ & 0.890 & $17.34 \pm 0.02$ & 54010.4622 & $0.53 \pm 0.04 $ & $-0.08 \pm 0.03$ & $17.29 \pm 0.04$ \\
		$J$ & 1.228 & $16.47 \pm 0.13$ & 51455.2390 & $0.86 \pm 0.06$ & $0.02\pm 0.09$ & $16.49 \pm 0.16 $ \\
		$H$ & 1.639 & $16.10 \pm 0.22$ & 51455.2390 & $0.86 \pm 0.06$ & $0.02\pm 0.09$ & $16.12 \pm 0.24 $\\
		$K_s$ & 2.152 & $>15.63$ & 51455.2390 & $0.86 \pm 0.06$ & $0.02\pm 0.09$ & $>15.65$ \\
		$W_1$ & 3.292 & $16.25\pm 0.06$ & $\sim$55400 & -- & -- & $16.25\pm 0.06$ \\
		$W_2$ & 4.542 & $16.71 \pm 0.25$ & $\sim$55400 & --& -- & $16.71 \pm 0.25$ \\
		$W_3$ & 10.302 & $>12.50$ & $\sim$55400 & -- & -- & $>12.50$ \\
		$W_4$ & 21.810 & $>9.15$ & $\sim$55400 & -- & -- & $>9.15$ \\

		\hline
	\end{tabular}
\end{table*}

This correction is complicated somewhat by the fact that the variability amplitude is likely wavelength dependent. Based on our fit to the spectrum, the phased light curves in Figure~\ref{fig:ztf_lcs}, and an initial fit to the SED without phase corrections, we estimate that the $r$-band variability amplitude is an adequate approximation of the variability in the optical and near-infrared bands. However, the more massive WD likely dominates the light in the UV, so we estimate that the variability amplitude is reduced by 30\% in the NUV  and is negligible in the FUV. The ``corrected mag'' column of Table~\ref{tab:sed} lists the magnitudes we assume in modeling the SED. In addition to corrections due to phase-variability, we also apply the empirical corrections from \citet{Eisenstein2006} to bring SDSS magnitudes to the AB magnitude system: $u_{\rm AB} = u_{\rm SDSS}-0.04$, $i_{\rm AB} = i_{\rm SDSS}+0.015$, and $z_{\rm AB} = z_{\rm SDSS}+0.03$.

\subsubsection{X-ray upper limit}
\label{sec:xray}
No X-ray source has been detected at the location of J0140, but the ROSAT survey \citep{Voges1999} obtained a 493 second exposure covering the object's coordinates. Using the ESA upper limit server,\footnote{See http://xmmuls.esac.esa.int/upperlimitserver/. We model the spectrum as a power law with slope 2 and assume an absorption column density $N_{\rm H} = 3\times 10^{20} \rm cm^{-2}$.} we obtain a 3-sigma upper limit of $2.5\times 10^{-13} \rm erg\,s^{-1}\,cm^{-2}$ in the 0.2-2 keV range. At a distance $d=1.5\,$kpc (Section~\ref{sec:constraints}), this corresponds to a luminosity $L_X \leq 0.018 L_{\odot}$ in the 0.2-2 keV range.

\begin{figure*}
    \centering
    \includegraphics[width=\textwidth]{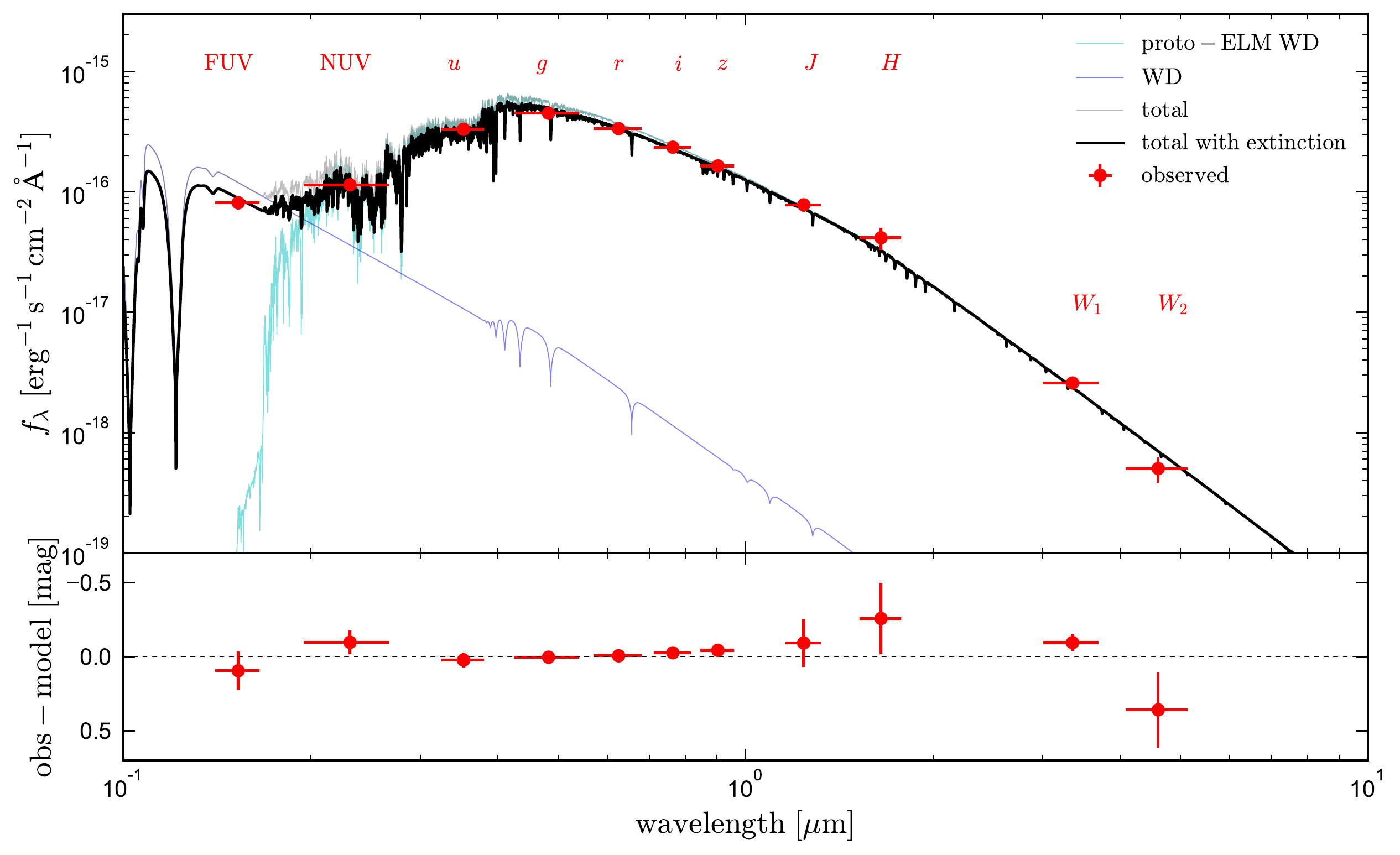}
    \caption{Spectral energy distribution of J0140. Red points with error bars show observed broadband fluxes from GALEX, SDSS, 2MASS, and WISE. Cyan shows a model spectrum for the proto-WD with $T_{\rm eff}= 6825\,\rm K$, $\log(g/{\rm cm\,s^{-2}})=4.7$, and angular diameter $\Theta = 2.05\,\mu \rm as$. Blue line shows a model for the more massive WD. Gray and black lines show the total model spectrum with and without extinction. Most of the SED is dominated by the proto-WD. However, it contributes negligibly in the FUV, where the SED is likely dominated by the more massive WD. Bottom panel shows the difference between the observed bandpass-integrated magnitudes and those predicted by the total SED model.}
    \label{fig:SED}
\end{figure*}

\subsubsection{Angular diameter and radius of the donor}
\label{sec:angular diameter}

%The SED provides a joint constraint on the effective temperatures and angular diameters, $\Theta = 2R/d$, of both the proto-WD and more massive WD.

We fit the SED for the effective temperatures and angular diameters, $\Theta = 2R/d$, of both components, as well as the reddening, $E(B-V)$. We predict the bandpass-averaged magnitudes of the proto-WD using synthetic photometry from MIST models \citep{Choi_2016}. We fix $\log g = 4.7$ (similar to the value implied by our fit in Section~\ref{sec:constraints}) since varying $\log g$ at fixed angular diameter has minimal effects on the SED. We calculate synthetic photometry for the proto-WD using atmospheric models from \citet{Koester2010} with $\log g = 8.5$. We assume a \citet{Cardelli_1989} extinction law with total-to-selective extinction ratio $R_V =3.1$. 

The temperature and angular diameter of the proto-WD are independently well-constrained by the SED: the wavelength at which the SED peaks constrains $T_{\rm eff}$, and the normalization in the Rayleigh–Jeans tail then constrains $\Theta$. For the more massive WD, $T_{\rm eff}$ and $\Theta$ are highly covariant, because the peak of its SED is likely blueward of the bluest bandpass observed. To break this degeneracy and constrain the more massive WD's temperature, we adopt a prior on its angular diameter, $\Theta_{\rm WD}/\mu {\rm as}\sim \mathcal{N}(0.055, 0.013)$. This is obtained from the {\it Gaia} distance and our dynamical constraint on the WD's mass (Section~\ref{sec:constraints}), assuming the mass-radius relation from \citet{Bedard2020}. We adopt a prior of $E(B-V)\sim \mathcal{N}(0.04, 0.02)$ based on the \citet{Green2019} 3D dust map, and flat priors on $T_{\rm eff,\,proto\,WD}$, $\Theta_{\rm proto\,WD}$, and $T_{\rm eff,\,WD}$.

We sample from the posterior using \texttt{emcee} \citep{emcee2013}, monitoring chains for convergence. The resulting constraints on the effective temperatures and angular diameters of both components are listed in Table~\ref{tab:system}, and the best-fit SED model is compared to the data in Figure~\ref{fig:SED}. When combined with the $\it Gaia$ parallax, the constraint on the angular diameter of the proto-WD translates to a constraint on its radius, $R_{{\rm proto\,WD}}=\left(0.33\pm0.09\right)R_{\odot}$.

If the proto-WD is assumed to fill its Roche lobe, as the light curve suggests it very nearly does (Section~\ref{sec:constraints}), then its mean density $\overline{\rho}=3M /\left (4\pi R^3\right)$ is given by
\begin{align}
    \label{eq:rhostar}
    \overline{\rho} \approx 107\,{\rm g\,cm^{-3}}\left(P_{\rm orb}/{\rm hr}\right)^{-2},
\end{align}
where $M$ and $R$ represent the mass and equivalent radius of the Roche-filling star \citep{Eggleton_1983}. The radius constraint from the SED then implies a proto-WD mass of $M_{\rm proto\,WD}=0.18^{+0.21}_{-0.11} M_{\odot}$.

The observed UV-to-IR SED is reproduced reasonably well. 
The SED of the proto-WD dominates in the optical and IR but drops off precipitously in the FUV. Under the assumption that the FUV emission is dominated by the more massive WD (without, for example, a significant contribution from accretion), we find a WD effective temperature $T_{\rm eff,\,WD}=19700\pm 1500$\,K. 

Under the assumption that the WD is accreting and has reached an equilibrium temperature for its accretion rate, its effective temperature can be related to the mass transfer rate \citep[e.g.][their Equation 2]{Townsley2009}. Taking $M_{\rm WD}=0.95 M_{\odot}$ (Section~\ref{sec:constraints}), the implied mass transfer rate is 
\begin{equation}
    \label{eq:mdot_teff}
    \dot{M} \lesssim 1.5\times 10^{-10} M_{\odot}\,\rm yr^{-1}.
\end{equation}
This is significantly lower than is found and predicted for most CVs at $P\approx 4$ hours, where most non-magnetic systems have $8\lesssim  - \log(\dot{M}/( M_{\odot}\,{\rm yr^{-1}}))\lesssim 9$. \citep[e.g.][]{Townsley2009, Kalomeni_2016}. Equation~\ref{eq:mdot_teff} is an upper limit, since the WD could remain hot from earlier accretion after a drop in the mass transfer rate.

\section{Parameter constraints}
\label{sec:constraints}
To constrain the physical parameters of the binary, we combine constraints from the ZTF light curves (Figure~\ref{fig:ztf_lcs}), Kast spectrum (Figure~\ref{fig:spectra}), multi-epoch RVs (Figure~\ref{fig:rvs}), photometric SED (Figure~\ref{fig:SED}), and the {\it Gaia} parallax. 

\subsection{Light curve model}
\label{sec:lc}
We model the ZTF $g-$ and $r-$band light curves simultaneously, using the binary light curve modeling software \texttt{Phoebe} \citep{prsa2005, prsa2016, horvat2018}, which builds on the \citet{Wilson1971} code. \texttt{Phoebe} models the surfaces of both components using a deformable mesh, self-consistently accounting for ellipsoidal variation, reflection, and  eclipses. The atmosphere of the donor is modeled using \texttt{Phoenix} model atmospheres \citep{Husser2013}  with $T_{\rm eff}=6800$\,K, and its limb darkening coefficients are calculated self-consistently from the atmosphere models. The WD, which contributes about 1\% of the light in the optical, is modeled with a blackbody atmosphere.  We assume it follows the mass-radius relation from \citet{Bedard2020} and has an effective temperature of 20,000 K. We fix the eccentricity to 0.

We leave the gravity darkening exponent $\beta_1$, where $T_{\rm eff}^4 \propto g^{\beta_1}$ \citep{vonZeipel_1924}, as a free parameter. It is common to assume $\beta_1=0.32$ for stars with significant convective envelopes ($T_{\rm eff} \lesssim 6300\,\rm K$), and $\beta_1=1$ for stars with radiative envelopes ($T_{\rm eff} \gtrsim 8000\,\rm K$; e.g., \citealt{Lucy1967}). J0140 is likely between these two regimes, with a thin convective envelope. Observational determinations of $\beta_1$ in this temperature range  are typically $0.2 \lesssim \beta_1 \lesssim 0.6$ \citep[e.g.][]{Claret2015}. We adopt a prior $\beta_1 \sim \mathcal{N}(0.32, 0.1)$.  

The light curve provides a joint constraint on the inclination and Roche lobe filling factor, $R_{\rm proto\,WD}/R_L$. We simultaneously fit the $g-$ and $r-$band light curves for these parameters,\footnote{In practice, this is accomplished by fixing the masses of the two components and leaving the radius of the proto-WD free. This is a sensible choice because, in the absence of detectable eclipses, the light curve shape depends just on the Roche lobe filling factor (which depends on the star's mean density) and on the limb darkening and gravity darkening coefficients, not on the absolute masses and radii.} as well as the intrinsic scatter in the light curves. Like the ``jitter'' term used in fitting the RVs, the intrinsic scatter term is added in quadrature to the reported photometric uncertainties; it essentially forces the reduced $\chi^2$ of the light curve fit to be 1.  We sample the posterior with \texttt{emcee}; our priors and constraints are listed in the first section of Table~\ref{tab:constraints}.

\begin{figure*}
    \centering
    \includegraphics[width=\textwidth]{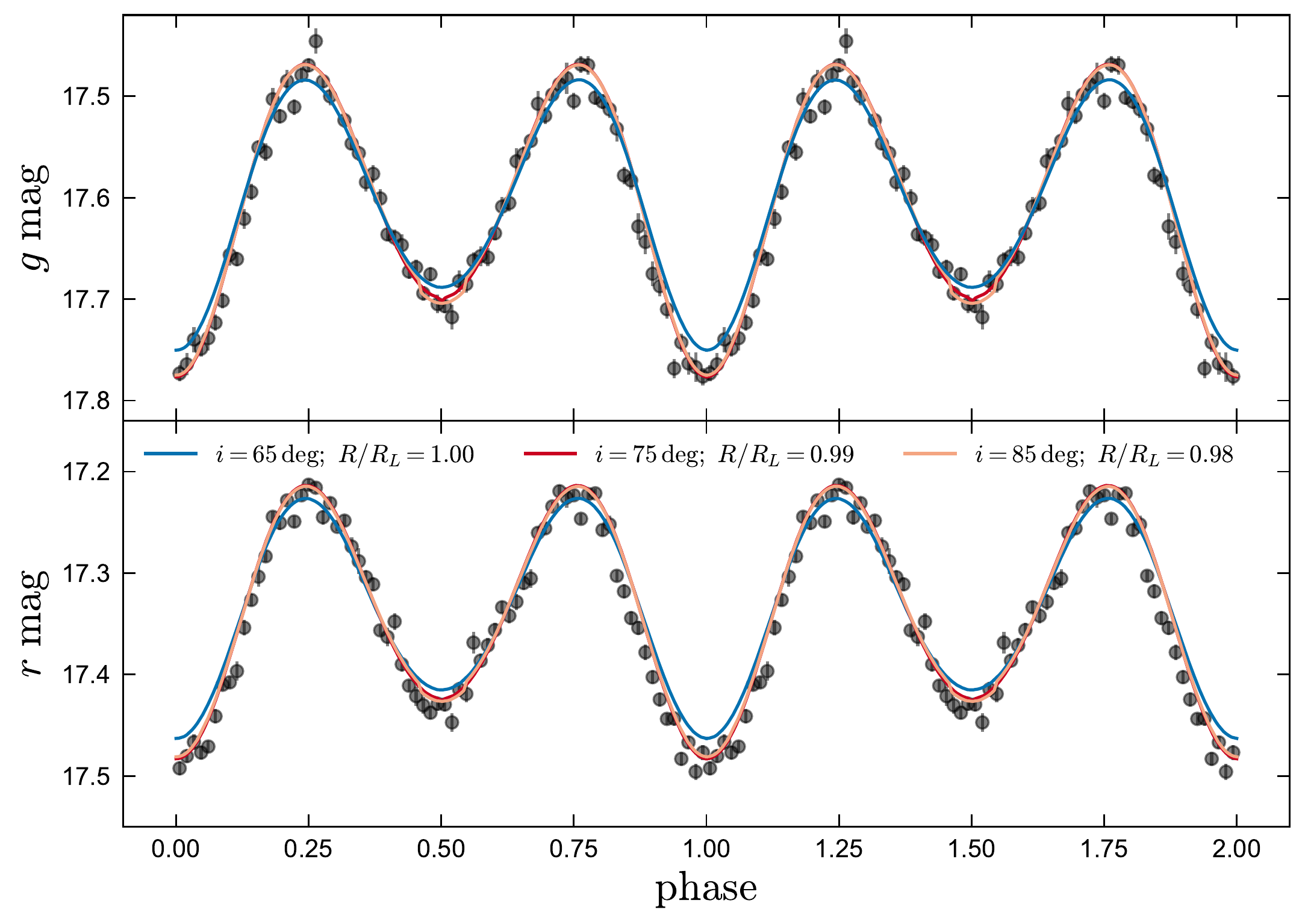}
    \caption{Best-fit \texttt{Phoebe} $g$ (top) and $r$ (bottom) light curves when the inclination is fixed to 60, 75, and 85 degrees. Models are compared to the phase-averaged ZTF data, with error bars including only random photometric uncertainties. Models are labeled by the ratio of the proto-WD equivalent radius to the equivalent Roche lobe radius. The best-fit models with $i=75$ and $i=85$ deg are nearly indistinguishable. Models with $i \gtrsim 75$ deg include a shallow eclipse at phase 0.5, when the massive WD is eclipsed, but the observed light curve is as yet too noisy to confirm or rule out this feature. The model with $i=65$ deg produces too-low ellipsoidal variability amplitude, even when the proto-WD fully fills its Roche lobe. Our fit to the light curve yields $i > 70$ deg and $R_{\rm proto\,WD}/R_{L} > 0.98$.}
    \label{fig:light_curve_inclination}
\end{figure*}

Figure~\ref{fig:light_curve_inclination} compares our best-fit \texttt{Phoebe} models to the observed light curves. For illustrative purposes, we show the best-fit models when the inclination is fixed at three different values. For all inclinations $i \gtrsim 70$ degrees, equally good fits to the light curve can be obtained by decreasing the Roche lobe filling factor while increasing the inclination. For models with $i\geq 75$ degrees, the eclipse of the massive WD is visible at $\phi = 0.5$. However, the depth of the predicted eclipse is less than 1\% in both bands, and the observed light curves cannot confidently distinguish between models that do and do not include eclipses. For inclinations $i < 70$ degrees, the predicted amplitude of ellipsoidal variability is weaker than what is observed, even when the proto-WD completely fills its Roche lobe. The blue curve in Figure~\ref{fig:light_curve_inclination} shows the best-fit model with $i=65$ deg, which is confidently ruled out by the data. 

The \texttt{Phoebe} models provide reasonable but not perfect fits to the light curve. Particularly in the $r-$ band, there are coherent residuals of $\approx 0.02$ mag at phases $\phi=0.75-1.25$. This could be the result of systematics in the model light curves (e.g., an inaccurate limb darkening law) or additional components not included in the model (e.g. an accretion disk or hot spot around the more massive WD, or spots on the surface of the proto-WD). Because the model does not perfectly describe the data, the formal uncertainties on light curve fitting parameters may be underestimated. We therefore do not fit the light curve and other observables simultaneously. Instead, we adopt priors on the inclination and Roche lobe filling factor of the proto-WD that are based on the light curve in our combined model, as described below.  

\subsection{Combined model}
\label{sec:combined}

We now combine constraints on the angular diameter, parallax, orbital velocity, inclination, and Roche lobe filling factor. The free parameters of the fit are the mass and radius of the proto-WD, orbital inclination, distance, and the mass of the more massive WD. The full likelihood function is: 
\begin{align*}
    \ln L = &-\frac{1}{2}\frac{\left[\left(R_{{\rm proto\,WD}}/R_{L}\right)_{{\rm model}}-\left(R_{{\rm proto\,WD}}/R_{L}\right)_{{\rm obs}}\right]^{2}}{\sigma_{R_{{\rm proto\,WD}}/R_{L}}^{2}}\\ &-\frac{1}{2}\frac{\left(K_{{\rm model}}-K_{{\rm obs}}\right)^{2}}{\sigma_{K,{\rm obs}}^{2}}-\frac{1}{2}\frac{\left(\Theta_{{\rm model}}-\Theta_{{\rm obs}}\right)^{2}}{\sigma_{\Theta,\,{\rm obs}}^{2}} \\
    &-\frac{1}{2}\frac{\left(\varpi_{{\rm model}}-\varpi_{{\rm obs}}\right)^{2}}{\sigma_{\varpi,\,{\rm obs}}^{2}}
\end{align*}
 The four terms compare the proto-WD's predicted Roche lobe filling factor, RV semi-amplitude, and angular diameter, and the system parallax, to the observed constraints. The proto-WD's  predicted Roche lobe equivalent radius is 
 \begin{equation}
    \label{eq:roche_filling}
    R_L \approx\frac{0.49q^{2/3}}{0.6q^{2/3}+\ln\left(1+q^{1/3}\right)}\left(\frac{P_{\rm orb}^{2}G\left(M_{{\rm proto\,WD}}+M_{{\rm WD}}\right)}{4\pi^{2}}\right)^{1/3},
\end{equation}
where $q=M_{\rm proto\,WD}/M_{\rm WD}$, and we use the fitting function from \citet{Eggleton_1983}. From the light curve fit, we take $\left(R_{\rm proto\,WD}/R_L\right)_{\rm obs} = 0.99$ and $\sigma_{R_{{\rm proto\,WD}}/R_{L}}^{2}=0.01$.
 
The predicted RV semi-amplitude of the proto-WD is
\begin{align}
    \label{eq:Kpred}
    K_{{\rm model}}&=\frac{\left[2\pi G\left(M_{{\rm proto\,WD}}+M_{{\rm WD}}\right)/P_{\rm orb}\right]^{1/3}}{\left(1+M_{{\rm proto\,WD}}/M_{{\rm WD}}\right)}\sin i,
\end{align}
the model parallax is $\varpi_{\rm model}=1/d$, and the angular diameter is $\Theta_{{\rm model}}=2R/d$. The observational constraints on these parameters are taken from Table~\ref{tab:system}.

Our adopted priors are listed in Table~\ref{tab:constraints}. We use flat priors on all parameters except distance. We take the distance prior from \citet{BailerJones2020}:
\begin{equation}
    \label{eq:distprior}
    P\left(d\right)=\begin{cases}
\frac{1}{\Gamma\left(\frac{\beta+1}{\alpha}\right)}\frac{\alpha}{L^{\beta+1}}d^{\beta}e^{-\left(d/L\right)^{\alpha}}, & {\rm if}\,d\geq0\\
0, & {\rm otherwise}
\end{cases}
\end{equation}
Here $L$, $\alpha$, and $\beta$ are free parameters that depend on sky position and are provided by \citet{BailerJones2020}. They were calculated by fitting the distance distribution of sources in a mock {\it Gaia} eDR3 catalog \citep{Rybizki2020}. For the location of J0140, $L=4.2\times 10^{-4}\,\rm pc$, $\alpha = 0.24$, and $\beta =  7.2$. The effects of the prior on our final constraints are weak; adopting a flat prior instead changes the best-fit distance and donor radius by only a few percent.

\begin{figure*}
    \centering
    \includegraphics[width=\textwidth]{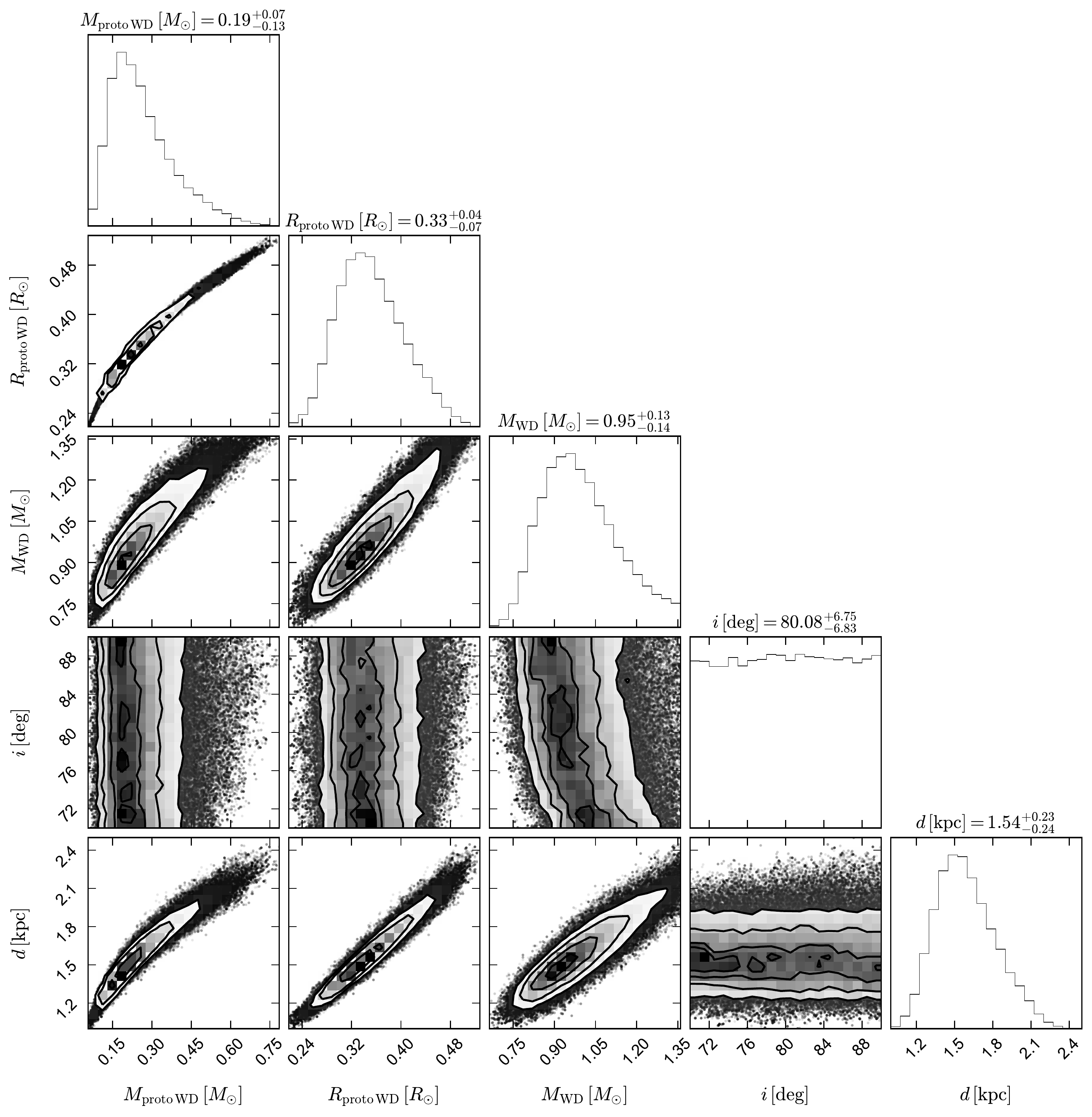}
    \caption{Parameter constraints from joint fitting of the light curve, SED, parallax, orbital solution, and donor spectrum. Marginalized constraints on each parameter are shown on the diagonal. }
    \label{fig:corner}
\end{figure*}

We sample the posterior distribution using \texttt{emcee} \citep{emcee2013}. We draw $ 4\times 10^5$ samples using 64 walkers, after a burn-in period of 2000 steps per walker. We inspect the chains to verify convergence. The resulting constraints are listed in the lower section of Table~\ref{tab:constraints} and shown in Figure~\ref{fig:corner}. The proto-WD luminosity implied by our constraints on its radius and effective temperature is $L_{{\rm proto\,WD}}=0.23^{+0.09}_{-0.06} \,L_{\odot}$. The surface gravity implied by the mass and radius constraints is $\log\left[g/\left({\rm cm\,s^{-2}}\right)\right] = 4.74\pm 0.07$, consistent with the spectroscopic fit. We compare these parameter constraints to models in Section~\ref{sec:mesa}.

\begin{table*}
\centering
\caption{Constraints from the combined fit. Uncertainties are 1$\sigma$. We first fit the light curve, adopting minimally informative priors on the inclination and Roche lobe filling factor. We then use the constraints from the light curve (namely, $i > 70$ deg and $R_{\rm proto\,WD}/R_L = 0.99\pm 0.01$) as priors during the combined fit. Finally (bottom two rows), we report constraints obtained when we require the proto-WD's period and effective temperature to match the binary evolution models introduced in Section~\ref{sec:mesa}. }
\begin{tabular}{lllll}
Parameter & description & units & prior & constraints \\

\hline

\multicolumn{5}{l}{\bf{Light curve fit}}   \\ 
$R_{\rm proto\,WD}/R_{L}$ & Roche lobe filling factor & -- & $\mathcal{U}(0,1)$  & $0.99\pm 0.01$ \\
$i$ & inclination & [deg] & $\sin(i)$  & $83^{+5}_{-8}$ \\
$\beta_{1,\,\rm proto\,WD}$ & gravity darkening exponent & -- & $\mathcal{N}({0.32, 0.1})$  & $0.29^{+0.03}_{-0.05}$ \\
$\sigma_{\rm intr}$ & Light curve intrinsic scatter & [mag] & $\mathcal{U}({0, 1})$  & $0.025 \pm 0.002$ \\

\hline
\multicolumn{5}{l}{\bf{Combined fit}}   \\ 
$R_{\rm proto\,\,WD}$ & Proto-WD radius & [$R_{\odot}$] & $\mathcal{U}({0.01, 1})$  & $0.33^{+0.07}_{-0.04}$ \\
$M_{\rm proto\,\,WD}$ & Proto-WD mass & [$M_{\odot}$] & $\mathcal{U}({0.01, 1})$  &  $0.19^{+0.13}_{-0.07}$  \\
$M_{\rm WD}$ & White dwarf mass & [$M_{\odot}$] & $\mathcal{U}({0.5, 1.36})$  & $0.95^{+0.14}_{-0.13}$ \\
$i$ & inclination & [deg] & $\mathcal{U}(70,90)$  & $80^{+7}_{-7}$ \\
$d$ & distance & [kpc] & Equation~\ref{eq:distprior}  & $1.54^{+0.24}_{-0.23}$ \\

\hline
\multicolumn{5}{l}{\bf{Combined fit including evolutionary models}}   \\ 
$R_{\rm proto\,\,WD}$ & Proto-WD radius & [$R_{\odot}$] & $\mathcal{U}({0.01, 1})$  & $0.29\pm 0.01$ \\
$M_{\rm proto\,\,WD}$ & Proto-WD mass & [$M_{\odot}$] & $\mathcal{U}({0.01, 1})$  &  $0.15\pm 0.01$  \\

\hline
\end{tabular}
\begin{flushleft}

\label{tab:constraints}
\end{flushleft}
\end{table*}

\section{Outburst frequency limits}
\label{sec:no_outburst}
If J0140 is still undergoing mass transfer, one might expect it to periodically exhibit dwarf nova-like outbursts, which usually occur in CVs due to disk instability \citep[e.g.][]{Hameury_2020}. The light curve of J0140 does not show any convincing evidence of such outbursts (Figure~\ref{fig:full_lc}). Here we use this fact to constrain an upper limit on the outburst frequency.

To determine whether an outburst would have been detected, we require an estimate of the outburst duration and amplitude. For normal CVs, the duration of typical outbursts is correlated with orbital period \citep[e.g.][]{Szkody_1984, Warner_2003}; for a period of 3.81 hours, the typical width at half-maximum, $\Delta T_{0.5}$, is $\approx 2.5$ days. Similarly, the absolute magnitude of dwarf nova disks during outburst is correlated with orbital period; for a period of 3.81 hours, $M_{V,\rm max}\approx 4.75$ \citep{Warner_2003}. For J0141, this would correspond to a $\approx 2.5$ mag brightening during outburst. It is of course not obvious that the outburst duration and amplitude of typical CVs will be a good proxy for an unusual system like J0140. 

\begin{figure}
    \centering
    \includegraphics[width=\columnwidth]{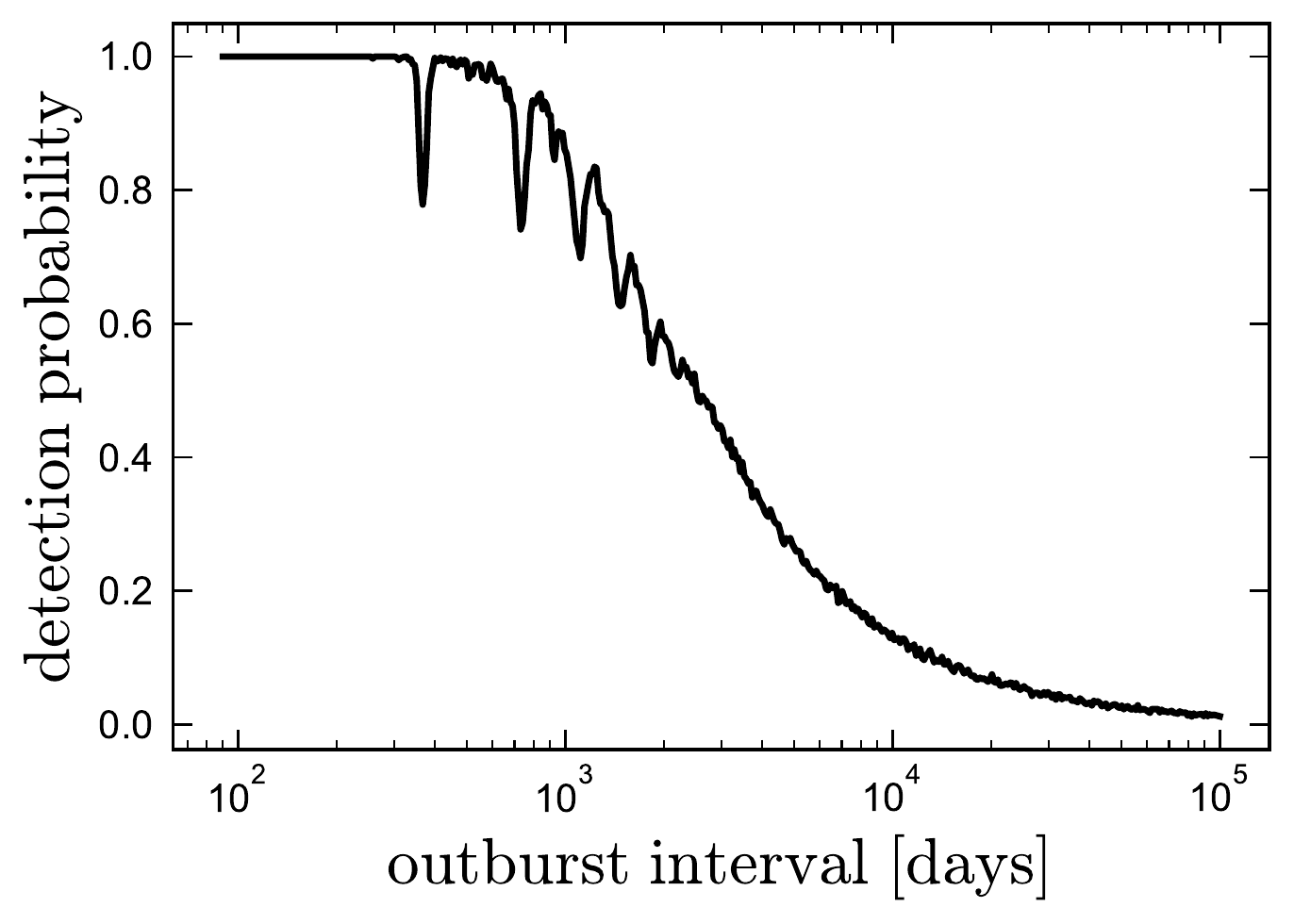}
    \caption{Probability that at least one outburst would have been detected by now if the true outburst interval were a given value, based on the light curve in Figure~\ref{fig:full_lc}. Sharp dips occur at multiples of 1 year, since J0140 is not observable for a few months each year; they would not be present if outbursts were not modeled as strictly periodic. The data rule out an outburst interval $T_n < 10^3$ days with 90\% probability.}
    \label{fig:outburst_detec_prob}
\end{figure}

To estimate the probability that outbursts with interval $T_n$ would have been detected by now, we perform Monte Carlo simulations in which we inject outbursts into the observed light curve (Figure~\ref{fig:full_lc}). For simplicity, we model each outburst as a tophat increase in brightness of 2.5 mag that lasts 2.5 days. If the magnitude during outburst exceeds an observed magnitude or upper limit, we conclude that it would have been detected. We model the bursts as strictly periodic, with uniformly distributed orbital phase. For each $T_n$, we perform 1000 trials and record the fraction of trials that result in at least one burst being detected. This fraction is plotted in Figure~\ref{fig:outburst_detec_prob}.

For all $T_n \lesssim 1$ year, the detection efficiency is 100\%. There are dips in the detection efficiency for $T_n$ values that are integer multiples of 1 year. This is because J0140 is not observable for a few months each year, so if bursts were phased such that they occur only during these gaps, they would not be detected. Real dwarf nova outbursts are generally not completely periodic, so the detection efficiency is likely smoother than modeled in this experiment. Figure~\ref{fig:outburst_detec_prob} shows that the probability of detecting at least one burst is greater than 90\% for $T_n \lesssim 10^3$ days. The detection efficiency falls below 50\% for $T_n \gtrsim 2500$ days.

\section{Evolutionary history and comparison to known CVs and ELM WDs}
\label{sec:mesa}

\begin{figure*}
    \centering
    \includegraphics[width=\textwidth]{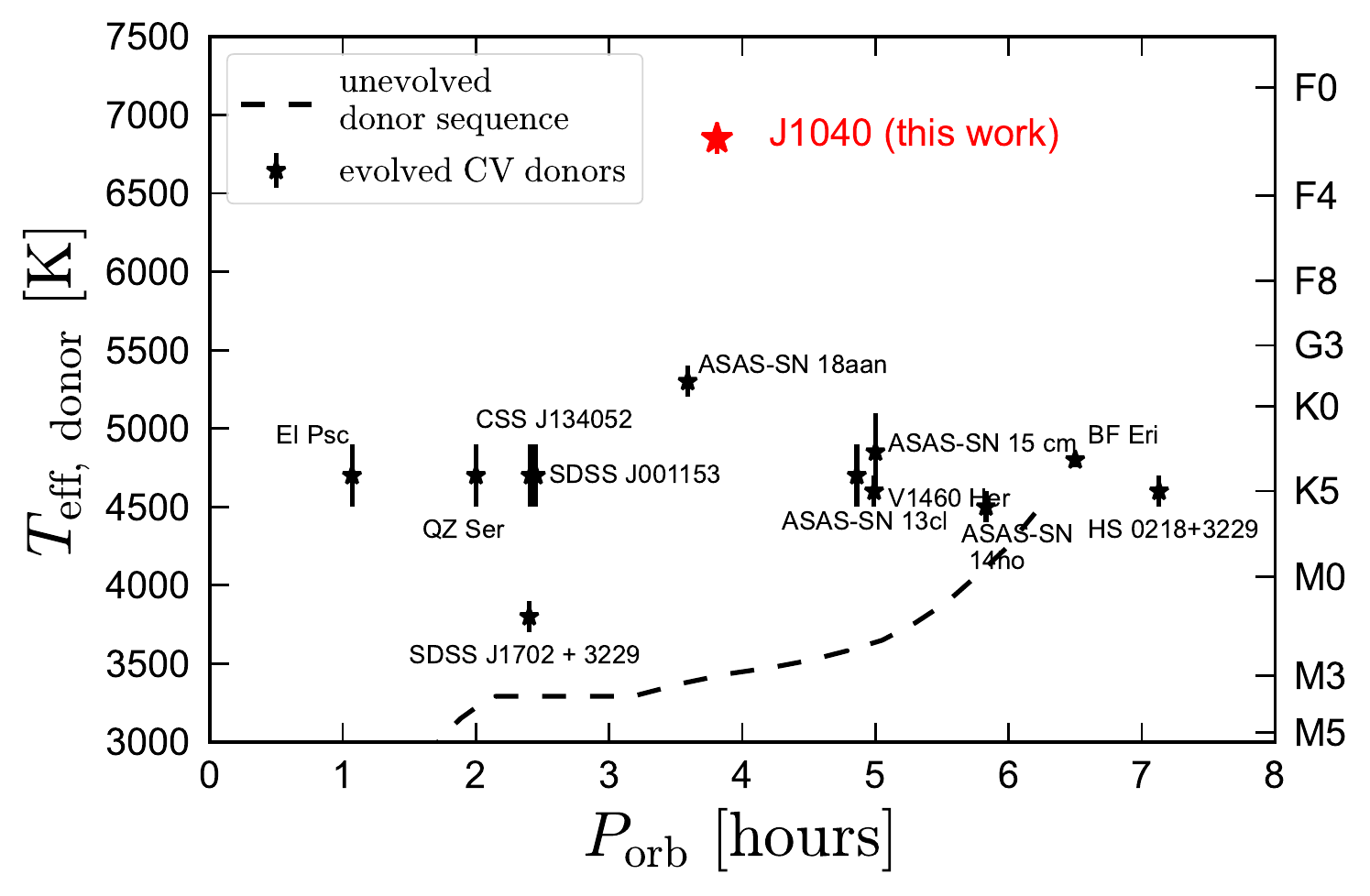}
    \caption{Effective temperatures and orbital periods of J0140 and known CV donors that have been proposed to have evolved donors (see Section~\ref{sec:mesa} for references). All these systems have hotter donors than predicted by the semi-empirical evolutionary sequence for unevolved donors (dashed line), which we take from \citet{Knigge_2006}. A majority of known CVs have temperatures close to this sequence. J0140 is considerably hotter than any previously characterized system at comparable period. }
    \label{fig:other_cvs}
\end{figure*}

If there is still ongoing mass transfer in J0140, then the proto-WD is an unusually evolved CV donor. If mass transfer was recently terminated, it is a just-detached ELM WD. We thus compare the system to known evolved CV secondaries and ELM WDs.

In Figure~\ref{fig:other_cvs}, we compare the proto-WD to CV donors from the literature that have been proposed to be evolved. These are QZ Ser \citep{Thorstensen_2002b}, EI Psc (also called 1RXS J232953.9+062814; \citealt{Thorstensen_2002}), SDSS J170213.26 + 322954.1 \citep{Littlefair_2006}, BF Eridani \citep{Neustroev2008}, CSS J134052.0 + 151341 \citep{Thorstensen_2013}, HS 0218+3229 \citep{RodriguezGil2009}, SDSS J001153.08-064739.2 \citep{Rebassa_2014},  ASAS-SN 13cl \citep{Thorstensen2015}, ASAS-SN 15cm \citep{Thorstensen2016}, ASAS-SN 14ho \citep{Gasque2019}, V1460 Her \citep{Ashley2020}, and ASAS-SN 18aan \citep{Wakamatsu2021}.

Most of these systems contain K star donors ($T_{\rm eff}=4500-5000$ K), which are warmer than is predicted for unevolved CV donors (dashed line; \citealt{Knigge_2006}). We note that evolved donors are common at long periods ($P_{\rm orb}\gtrsim 6$ hours), because only evolved or unusually massive donors are large enough to already be Roche-filling at these periods. There are plenty of evolved donors with longer periods that are not shown in Figure~\ref{fig:other_cvs}; some of these are shown in Figure~\ref{fig:evol_mesa}. However, evolved systems are rare at $P_{\rm orb}\lesssim 6$ hours, and the CVs shown in Figure~\ref{fig:other_cvs} represent a near-complete inventory of currently known systems with anomalously hot donors.  J0140 is significantly hotter than other known CVs at comparable periods. If it is still mass-tranferring, it is the most extreme evolved CV discovered to date. 

\subsection{MESA models}
\label{sec:mesa_mdoels}
To understand the evolutionary history of J0140 and similar systems, we calculated a grid of binary evolution models using Modules for Experimentation in Stellar Astrophysics \citep[MESA, version 12778;][]{Paxton_2011, Paxton_2013, Paxton_2015, Paxton_2018, Paxton_2019}. The binary capabilities of MESA are described in detail by \citet{Paxton_2015}. We follow the mass-losing star with detailed calculations and model the more massive WD as a point mass.  Roche lobe radii are computed using the fit of \citet{Eggleton_1983}. Mass transfer rates during Roche lobe overflow are determined following the prescription of \citet{Kolb_1990}. The orbital separation evolves such that the total angular momentum is conserved when mass is transferred to the companion or lost, as described in \citet{Paxton_2015}. We model mass transfer as being fully non-conservative, with the mass that is transferred from the donor to the WD eventually being lost from the vicinity of the WD, as described by \citet[][their ``$\beta$'' parameter]{Tauris_2006}. Although mass transfer in real CVs is not {\it instantaneously} non-conservative, the mass accreted by the WD is expected to be ejected from the system by classical novae explosions, which recur on a timescale that is short compared to the timescale on which the orbit evolves.

We initialize our calculations with a zero-age main sequence (ZAMS) star in a detached orbit with a point mass representing the WD. The orbit shrinks through gravitational wave radiation and magnetic braking. Magnetic braking is implemented following the prescription of \citet{Rappaport_1983}, which applies a torque,
\begin{align}
    \label{eq:Jdotmb}
    \tau_{{\rm mb}}=-6.8\times10^{-34}\left(\frac{M_{{\rm donor}}}{M_{\odot}}\right)\left(\frac{R_{{\rm donor}}}{R_{\odot}}\right)^{\gamma_{\rm mb}}\left(\frac{P}{1\,{\rm d}}\right)^{-3}{\rm dyn\,cm},
\end{align}
removing angular momentum from the binary. This expression is derived under the assumption that the binary is tidally locked and the angular momentum loss from the donor through winds is the same as that from an isolated single star with the same rotation velocity \citep[e.g.][]{Skumanich_1972, Verbunt_1981}. Comparison to the observed age-rotation period relation for solar-type stars suggests $\gamma_{\rm mb}=4$; \citet{Rappaport_1983} found $2\lesssim \gamma_{\rm mb}\lesssim 4$ when matching their evolutionary models to the period distribution of observed CVs. We adopt $\gamma_{\rm mb}=3$, but our results are only weakly sensitive to this choice.

Magnetic braking is turned off when the donor becomes fully convective, because at that time the magnetic field is expected to weaken, or at least significantly change its structure \citep{Spruit_1983, Donati_2009}. We also expect magnetic braking to become inefficient when the donor loses its convective envelope \citep[e.g.][]{Kraft_1967}. This is not implemented in the default distribution of MESA. Following \citet{Podsiadlowski_2002}, we assume that the magnetic braking torque follows Equation~\ref{eq:Jdotmb} as long as $q_{\rm conv}$, the mass fraction of the donor's convective envelope, exceeds 0.02. For donors with $q_{\rm conv} < 0.02$, we suppress the magnetic braking torque by multiplying Equation~\ref{eq:Jdotmb} by a factor $\exp(1-0.02/q_{\rm conv})$. This effectively makes magnetic braking inefficient at $T_{\rm eff} \gtrsim 6000\,\rm K$.

The ``initial'' separations in our calculations represent the separation after the more massive WD has formed, presumably through an episode of common envelope evolution (CEE). The separation distribution at this stage is not well-constrained observationally or theoretically, and we do not attempt to predict exactly what initial binary configurations would produce a given separation after CEE. We instead simply model a grid of separations to see which post-CEE separations could produce the observed properties of J0140. 

Initializing the secondary on the ZAMS neglects its aging during the main-sequence lifetime of the massive WD's progenitor. This is a reasonable approximation for our purposes, because, in order for the system to have gone through CEE and emerged as a WD+MS binary with a short period, the initial mass ratio of the binary must have been relatively unequal \citep[e.g.][]{Soberman_1997}. If the massive WD's mass has not changed significantly from accretion or novae, then its current mass of $\approx 0.95 M_{\odot}$ implies an initial progenitor mass between 3.5 and 5.5\,$M_{\odot}$, depending on the adopted initial-final mass relation \citep[e.g.][]{Cummings2018, Elbadry2018}. This corresponds to a pre-WD lifetime of at most 300 Myr.

\begin{figure*}
    \centering
    \includegraphics[width=\textwidth]{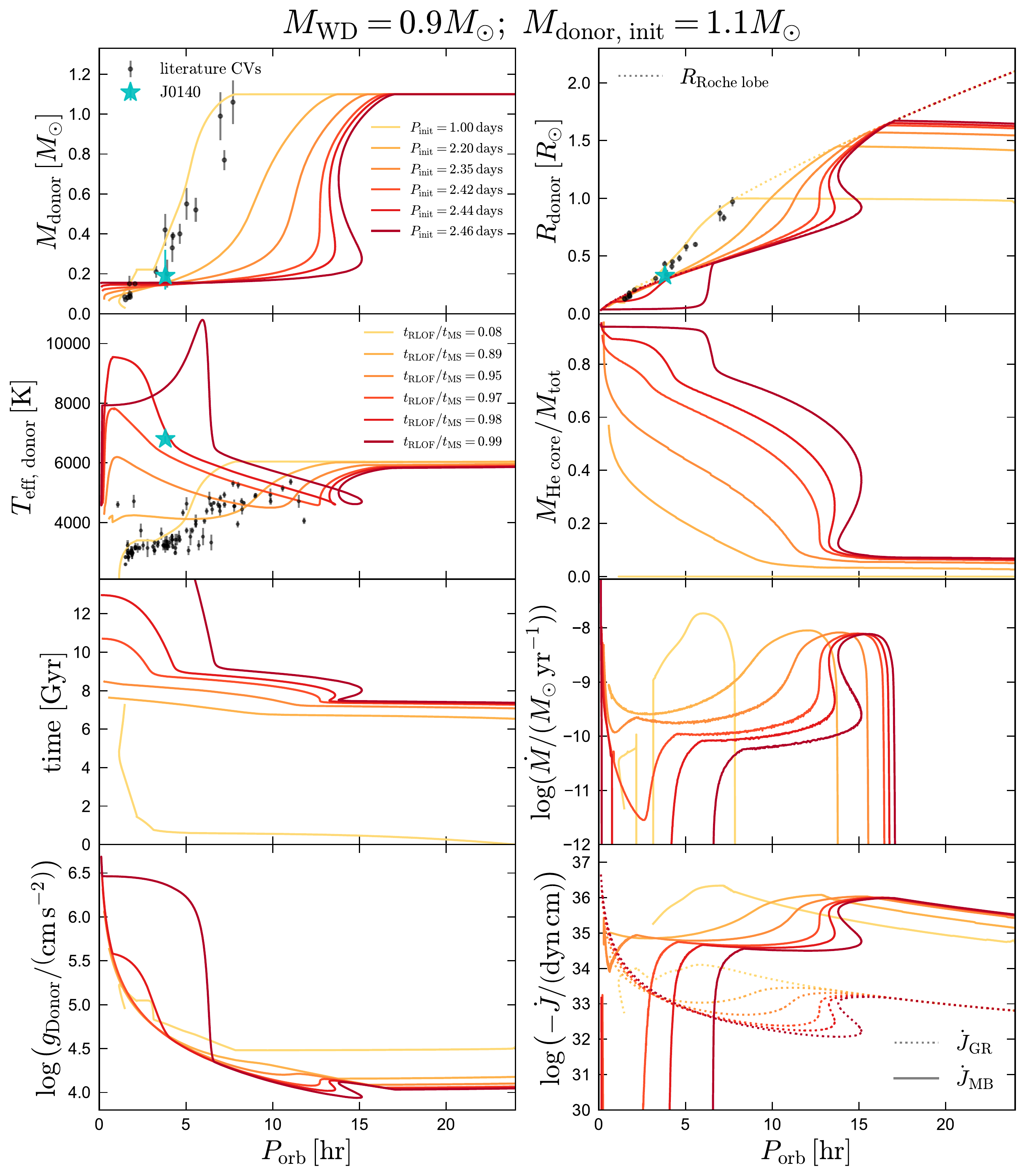}
    \caption{MESA binary models for CVs with evolved donors. We show 6 models, labeled by $P_{\rm init}$ (the initial period at which we initialize a detached WD+MS binary) and by the fraction of the donor's normal MS lifetime that has passed at the onset of RLOF.  Gray points with error bars show other CVs from the literature with well-measured masses, radii, and effective temperatures. The cyan point shows J0140. The yellow track with $P_{\rm init}=1$ day is representative of CVs in which the donor overflows its Roche lobe before undergoing significant nuclear evolution; it passes through the majority of the observed CV population. Darker tracks shows systems (with finely spaced periods) in which the donor overflows its Roche lobe increasingly near the end of the main sequence. By the time these systems reach $P_{\rm orb} \sim 4$ hours, they have significant helium cores with thin H-burning envelopes. At fixed period, they have higher $T_{\rm eff}$, lower masses and radii, and much lower mass transfer rates than ordinary CVs. J0140 is well described by such tracks. The most extreme evolved tracks become detached once they lose their convective envelopes and magnetic braking becomes inefficient (lower right); these become ELM WDs. }
    \label{fig:evol_mesa}
\end{figure*}

We ran MESA tracks with initial periods ranging from 0.5 to 3 days, WD masses ranging from 0.8 to 1.2 $M_{\odot}$, and initial donor masses ranging from 1 to 2.5 $M_{\odot}$. The qualitative behavior of the models depends little on the assumed WD mass, so here we only analyze those with $M_{\rm WD} = 0.9 M_{\odot}$. Figure~\ref{fig:evol_mesa} shows the evolution of several tracks that all have initial donor masses of $1.1 M_{\odot}$ and a range of initial orbital periods; we discuss other donor masses in Section~\ref{sec:vary_mesa}. As a function of orbital period, we plot the donor mass, radius, and effective temperature, the mass of its helium core\footnote{Defined as the region in which the helium mass fraction is >99\%.}, time since the beginning of the calculation, the mass transfer rate, the surface gravity of the donor, and the binary's angular momentum loss due to magnetic braking and gravitational wave radiation.
We also plot the mass, radius, and temperature of donors in known CVs with well-characterized donor parameters.\footnote{Effective temperatures for 91 CVs are taken from \citet[][their Table 2]{Knigge_2006}. Only CVs with reliable spectroscopically determined spectral types are included. \citet{Knigge_2006} report spectral types, not effective temperatures; we convert them to $T_{\rm eff}$ using the conversion from \citet{deJager1987}. We also plot masses and radii of donors in 19 eclipsing CVs from \citet[][their Table 8]{Patterson2005}. We plot these, rather than larger compilations from elsewhere in the literature, because they contain masses and radii measured independently, without the assumption that the CVs follow a particular donor sequence.}

The set of initial periods shown in Figure~\ref{fig:evol_mesa} are chosen to highlight the behavior of ``normal'' CVs as well as evolved systems like J0140. The track with $P_{\rm init}= 1$ day displays behavior characteristic of ``normal'' CVs: the donor overflows its Roche lobe while still on the main sequence, before a helium core has formed. It grows smaller and cooler as the period shrinks, passing near most of the observed systems.\footnote{Even this model is somewhat warmer that the bulk of the observed population at short periods. This likely reflects the fact that the atmospheres of cool dwarfs are influenced by molecules, which are not included in the MESA calculations. More reliable effective temperatures for cool dwarfs can be obtained by changing the boundary condition where model atmospheres are stitched to interior models \citep[see][their Figure 14]{Choi_2016}. We refrain from doing so here, using the  \texttt{photosphere\_tables} atmosphere boundary in all calculations. } At a period of about 3 hours, the donors become fully convective and magnetic braking ceases, slowing the orbital decay. At this point, the donor temporarily detaches from its Roche lobe and returns to thermal equilibrium as it passes through the period gap, before reestablishing Roche contact at $P_{\rm orb}\sim 2$ hours. The donor continues to shrink until the period minimum, where it becomes degenerate. The same qualitative evolutionary scenario applies to all models that overflow their Roche lobes while on the main sequence, including models with higher- and lower-mass secondaries. For the choice of $M_{\rm WD}$ and $M_{\rm donor,\,init}$ shown in Figure~\ref{fig:evol_mesa}, all systems with initial periods between about 0.4 and 2 days follow similar evolution to this track. 

Along with their initial periods, we label the tracks in Figure~\ref{fig:evol_mesa} by the fraction of the donor's main-sequence lifetime (if it were isolated; this is $t_{\rm MS} = 7.54$\,Gyr for the 1.1 $M_{\odot}$ donor shown in the figure) that has passed at the time of first Roche lobe overflow. We define the main-sequence lifetime as the time until the helium core mass reaches the Schönberg–Chandrasekhar limit, which we take to be 0.08 times the total mass of the star. Models with $t_{\rm RLOF}/t_{\rm MS}\lesssim 0.8$ all behave similarly to the track shown with $t_{\rm RLOF}/t_{\rm MS} = 0.08$. 

Models with longer initial periods and $t_{\rm RLOF}/t_{\rm MS} \gtrsim 0.8$ behave qualitatively differently. These donors overflow their Roche lobes as a helium core is already forming. Initial RLOF occurs at a significantly longer period than in CVs with unevolved donors, because the donor's radius has increased by $\sim$60\% during its main-sequence evolution. 
Once most of the envelope has been lost -- over a period of several hundred Myr -- the effective temperature of the donor begins to increase. The most evolved models, which have the thinnest hydrogen envelopes, reach the highest temperatures. By the time they reach $P_{\rm orb}\sim 4$ hours, the evolved donors have low mass loss rates, $\dot{M}\lesssim 10^{-10}M_{\odot}\,\rm yr^{-1}$.

The mass transfer rate for the evolved models in Figure~\ref{fig:evol_mesa} is never particularly high, reaching at most $10^{-8} M_{\odot}\,\rm yr^{-1}$. It can, however, be significantly higher for higher initial donor masses (Section~\ref{sec:vary_mesa}). In the most evolved models shown, the orbital period does not decrease monotically with time, but increases slightly while the donor's mass is between $\approx 0.6$ and $\approx 0.3 M_{\odot}$. The orbital separation {\it does} decrease monotonically; the period increase reflects a rapid decrease in total mass. 

The lower mass transfer rate near $P_{\rm orb}=4$ hours in models with evolved donors is a consequence of (a) these donors' smaller masses and radii, which lead to weaker magnetic braking torques (Equation~\ref{eq:Jdotmb}), and (b) the fact that the mass in evolved donors is more centrally concentrated, so a fixed decrease in radius corresponds to a smaller decrease in mass. 

Once the donor reaches $T_{\rm eff} \gtrsim 6000$\,K and loses its convective envelope, magnetic braking ceases. This can be seen in the bottom panel of Figure~\ref{fig:evol_mesa}, which shows a sharp fall-off in $\dot{J}_{\rm MB}$ for the evolved models that reach $T_{\rm eff} \gtrsim 6000$\,K. The binary's orbital decay is then driven only by gravitational radiation. This leads to an immediate decrease in the mass transfer rate. In the two most evolved models, this causes the donor to detach from its Roche lobe and reestablish thermal equilibrium at a smaller radius. During this period, the donor would appear as a detached ELM WD. For the model that most closely matches the effective temperature of J0140, detachment occurs {\it just} at $P_{\rm orb}=3.8$ hours. Slightly less extreme models remain in Roche contact, either because magnetic braking never fully shuts off, or because the binary has reached  short enough periods that angular momentum loss through gravitational waves is efficient by the time magnetic braking becomes inefficient. 

It is important to note that there are few observational constraints on the efficiency of magnetic braking in stars like the proto-WD in J0140. 
The details of which models remain in Roche contact, which become detached, and when detachment occurs, depend sensitively on the adopted magnetic braking law. Characterization of similar CVs with hot and warm donors, particularly at $1 \lesssim P_{\rm orb}/{\rm hour} \lesssim 3$, will allow the efficiency of magnetic braking in hot stars to be better constrained. 

The evolutionary models also allow us to refine our estimate of the mass and radius of the proto WD: plausible models with evolved donors in Figure~\ref{fig:evol_mesa} have $M_{\rm proto\,WD}=0.15\pm 0.01\,M_{\odot}$ and $R_{\rm proto\,WD}=0.29\pm 0.01\,R_{\odot}$. Other published evolutionary tracks for ELM WDs imply consistent values \citep[e.g.][]{Sun_2018, Li2019}. Given the measured angular diameter, this would imply a distance $d=1.30\pm 0.05$\,kpc. These values are consistent with, but tighter than, our constraints that are independent of evolutionary models (Section~\ref{sec:constraints}).

\begin{figure*}
    \centering
    \includegraphics[width=\textwidth]{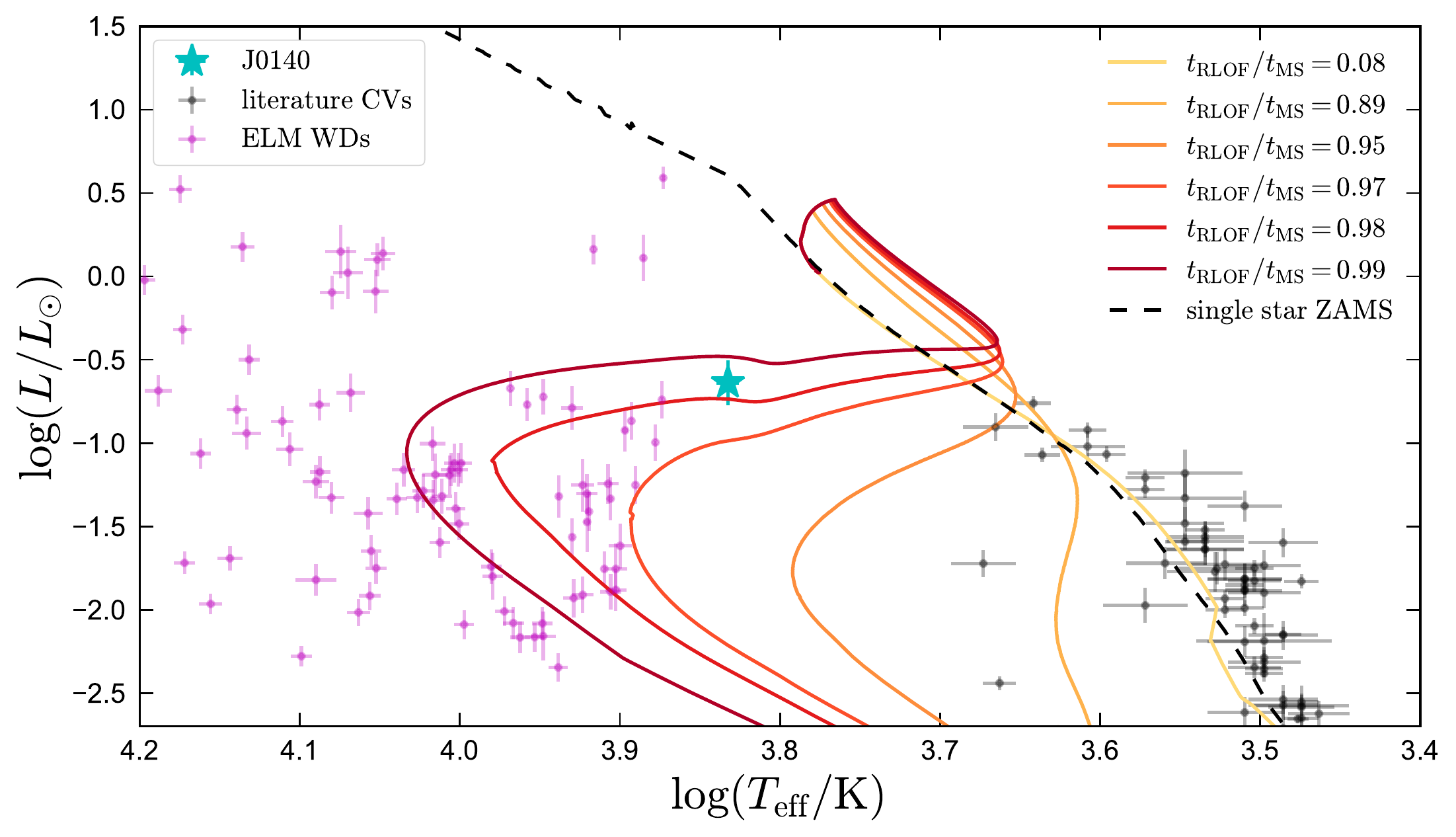}
    \caption{HR diagram. The same MESA tracks from Figure~\ref{fig:evol_mesa} are compared to other CV donors (gray points) and to extremely low mass (ELM) WDs. Most normal CV donors fall near the single-star main sequence, which also overlaps the yellow model with early RLOF. Tracks in which the donor overflows its Roche lobe near the end of its main-sequence evolution (orange and red) diverge from the main sequence toward higher $T_{\rm eff}$, approaching the ELM white dwarfs. The most extreme models become detached in this phase (Figure~\ref{fig:evol_mesa}). J0140 is likely transitioning between the two populations, evolving to higher temperature at near-constant luminosity. }
    \label{fig:hrd_mesa}
\end{figure*}

Figure~\ref{fig:hrd_mesa} shows the evolution of the same MESA models on the HR diagram. The unevolved CV track with $t_{\rm RLOF}/t_{\rm MS}=0.08$ closely follows the single-star main sequence (dashed black line). Most of the CVs in the sample to which we compare fall near this track, though they are on average offset slightly to lower $T_{\rm eff}$. A few literature systems (gray points) are significantly hotter than the single-star main sequence at their luminosity; these are CVs thought to host evolved donors. J0140 is significantly offset from these systems. The MESA tracks that pass near it suggest it will continue evolving toward higher $T_{\rm eff}$ at nearly fixed luminosity before it eventually cools. 

Figure~\ref{fig:hrd_mesa} also shows a sample of spectroscopically confirmed ELM WDs assembled by \citet{Pelisoli2019}. A majority of these objects were discovered by the ELM survey \citep{Brown2010}. We estimate the luminosity of these objects from their reported $T_{\rm eff}$, $\log g$, and mass.\footnote{That is, we calculate $L=L_{\odot}\left(R/R_{\odot}\right)^{2}\left(T_{{\rm eff}}/T_{{\rm eff},\odot}\right)^{4}$, with $\left(R/R_{\odot}\right)^{2}=G\left(M/M_{\odot}\right)\left(g/g_{\odot}\right)^{-1}$. Since no mass uncertainty is reported for most of these objects, we assume an uncertainty of $0.03 M_{\odot}$ when calculating the uncertainty in $L$.}  We note that selection effects for these objects and for CVs are very different, and we make no attempt to interpret the relative numbers of objects in the two populations. ELM WDs are found at even higher $T_{\rm eff}$ than J0140, and they are all detached. The two most extreme MESA tracks run through this population in temperature--luminosity space after detachment. This suggests that, whether or not it is currently mass transferring, J0140 will evolve to become a detached ELM WD. We note that a majority of the ELM WDs plotted have somewhat longer orbital periods than J0140. Such systems are thought to form if $t_{\rm RLOF}/t_{\rm MS} > 1$; i.e., if the donor's core has reached the Schönberg–Chandrasekhar limit before first RLOF. In this case, the donor radius and separation expand in response to mass loss, and the system never reaches short periods \citep[e.g.][]{Kalomeni_2016, Sun_2018}.

\subsection{Future evolution}
\label{sec:future}
\begin{figure}
    \centering
    \includegraphics[width=\columnwidth]{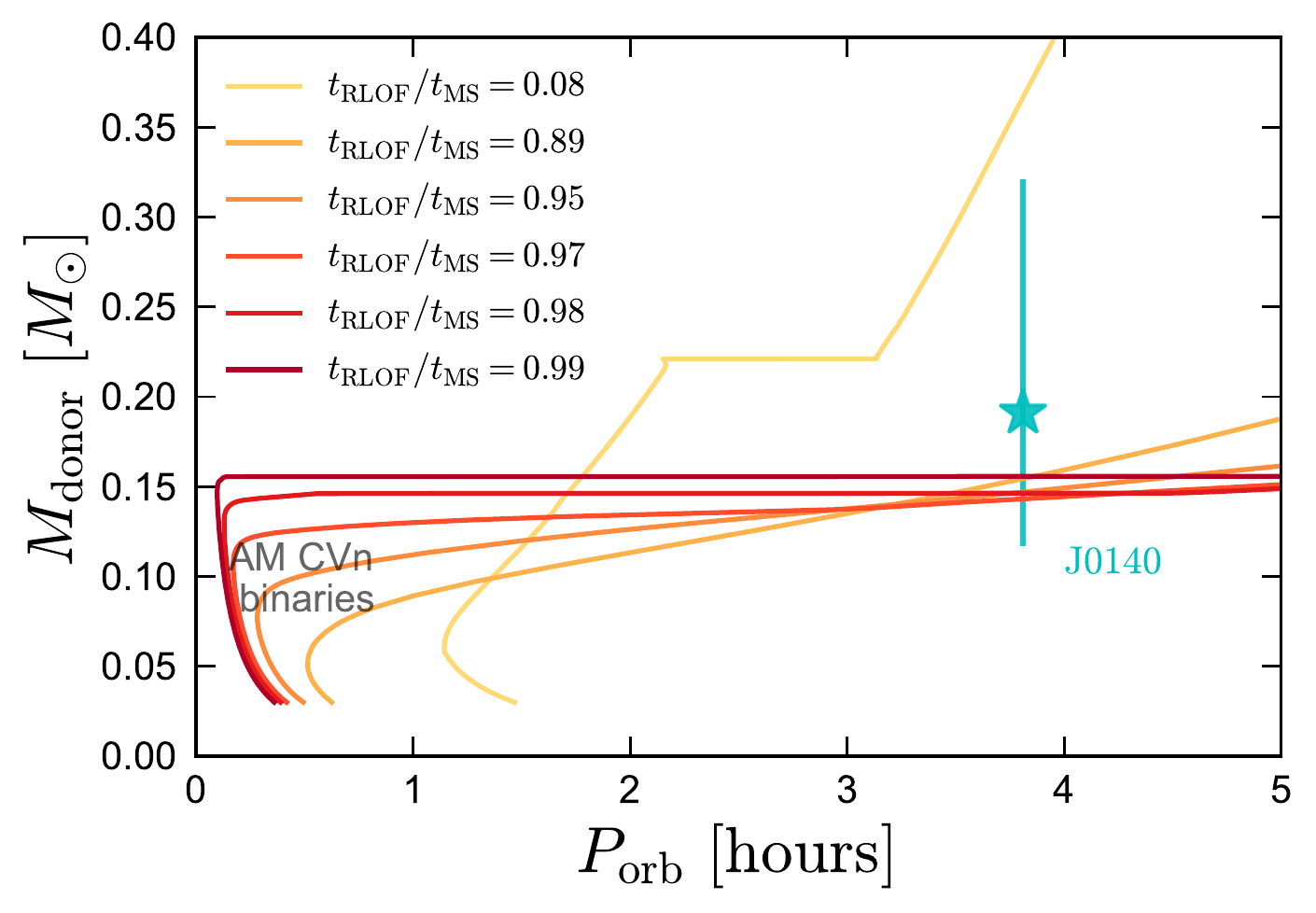}
    \caption{Future evolution of J0140. Colored tracks show the same MESA models from Figures~\ref{fig:evol_mesa} and~\ref{fig:hrd_mesa}. Normal CVs reach a period minimum at about 80 minutes, when they become degenerate. Systems with evolved donors are smaller when they become degenerate (due to higher mass and lower $Y_e$), allowing them to reach much shorter minimum periods. These models suggest that J0140 will reach a minimum period $P_{\rm orb}\lesssim 15$ minutes, when (if it avoids a merger) it will be classified as an AM CVn binary. }
    \label{fig:p_vs_m}
\end{figure}

Irrespective of whether angular momentum loss is driven by magnetic braking or gravitational waves, the orbit of J0140 will continue to decay.
Figure~\ref{fig:p_vs_m} shows the evolution of the MESA models from Figures~\ref{fig:evol_mesa} and~\ref{fig:hrd_mesa} at short periods. We run the calculations until the donor mass falls below $0.03\,M_{\odot}$, noting that in the most evolved model, this takes longer than the age of the Universe.

All models eventually reach a period minimum and then begin to widen again. The mass ratio is low enough that a merger is unlikely, at least under the simplified treatment of mass transfer and accretion onto the WD assumed here \citep[e.g.][]{Marsh2004}. The minimum period is much shorter for the evolved tracks, which reach minimum periods between 5 and 30 minutes, compared to $\approx 80$ minutes for the unevolved CV track. Mass transfer rates at the period minimum are high but not catastrophic, reaching $1.5\times 10^{-7} M_{\odot}\,\rm yr$ in the most evolved model. Models that reach short periods evolve through the period minimum relatively quickly; for example, the most evolved model spends only 20 Myr with $P_{\rm orb} < 20$ minutes. The period minimum of evolved CVs was explored extensively by \citet{Podsiadlowski_2003}, and the behavior of our models is qualitatively similar to their predictions.

The shorter period minimum for evolved donors can be understood as follows. The period minimum is set by the radius of the donor at the time when it becomes degenerate. Unevolved donors become degenerate at the hydrogen burning limit, $M_{\rm donor}\approx 0.08 M_{\odot}$. Due to modest departures from thermal equilibrium, their radius at this time is $R_{\rm donor}\approx 0.12 R_{\odot}$, about 20\% larger than the radius of a main-sequence star of the same mass. Equation~\ref{eq:rhostar} then predicts a period minimum of 77 minutes, close to what is found in the MESA models. 
Evolved donors have smaller radii. The radius of cold, non-relativistic degenerate objects scales as $R\sim M^{-1/3}Y_e^{5/3}$, where $Y_e$ is the mean number of electrons per nucleon. Taking $Y_e\approx 0.5$ for evolved donors that are mostly helium and $Y_e \approx 0.87$ for normal donors with Solar composition, the radius of a $0.16 M_{\odot}$ evolved donor (corresponding to the most evolved model in Figure~\ref{fig:p_vs_m}) at the period minimum is predicted to be $\approx 3.2$ times smaller than that of a $0.08 M_{\odot}$ hydrogen donor. This corresponds to a 64 times higher mean density and an 8 times shorter minimum period (Equation~\ref{eq:rhostar}). The actual minimum period is $\approx 13$ times shorter, likely a result of the fact that near the period minimum, the unevolved model is inflated by mass loss-driven departures from thermal equilibrium, while the evolved model (being detached just prior to the onset of mass transfer and the minimum orbital period) is not.

Near its period minimum, J0140 may appear as an ``AM-CVn'' binary, \citep[e.g.][]{Solheim_2010}.  An ``evolved CV'' channel for producing AM-CVn binaries is discussed in \citet{Podsiadlowski_2003}. Because a relatively fine-tuned scenario is required to produce an ultra-compact binary from a CV, most studies have concluded that the evolved CV channel likely only forms a small fraction of ultracompact binaries \citep{vandersluys2005, Solheim_2010, Goliasch_2015}. Nevertheless, all plausible models we construct for J0140 eventually reach short periods and either remain mass-transferring or resume mass transfer after the period minimum.

Other possible formation channels for AM-CVn binaries include a WD+WD channel \citep{Tutukov_1979} and a He-star-WD channel \citep{Iben_1991, Brooks_2015}; which channel dominates is not known. If J0140 becomes detached before the period minimum, it might be counted instead as contributing to the WD+WD channel, for which a few detached progenitor systems have been identified \citet{Kilic2014}. If J0140 becomes detached, it will evolve to become very similar to these systems. Near the period minimum, the system would then resemble some of the ultracompact WD binaries recently discovered by ZTF \citep[e.g.][]{Burdge2019, Burdge2020}.

The fate of the system after the period minimum is unclear. If a merger is avoided, accretion of helium onto the surface of the more massive WD will lead to unstable thermonuclear flashes, analogous to classical novae on normal CVs. 
These flashes are predicted to become less frequent and more energetic as the accretion rate decreases, eventually culminating in a ``final flash'' that may fuse radioactive elements and could be observed as a rapidly-evolving transient (a ``.Ia'' supernova; \citealt{Bildsten_2007}). Whether a detonation occurs in this scenario depends on the time-dependence of the accretion \citep{Piersanti2015}.

An alternative possibility is that, following classical nova-like flashes on the surface of the accreting WD, dynamical friction will shrink the orbit and cause the binary to merge \citep[][]{Shen_2015}. The outcome of such a merger would depend on the masses of the components. If the accreting WD has a carbon-oxygen core and the total mass exceeds $\approx 1.2 M_{\odot}$, then a SNe Ia may be triggered through surface He detonation and subsequent C detonation in the core by  a converging shock \citep[e.g.][]{Livne_1990, Pakmor_2013, Shen_2014}. For lower total masses (which are more likely, given our constraints), the merger might form an 
R Coronae Borealis star \citep[e.g.][]{Clayton2012}.

\subsection{Varying model parameters}
\label{sec:vary_mesa}
Here we briefly summarize how the evolution of the MESA models changes when the problem setup is varied. 
\begin{itemize}
    \item {\it Shorter or longer initial periods:} Shorter initial periods behave basically identically to the track with $P_{\rm init} = 1$ day. In binaries with longer initial periods, the donor overflows its Roche lobe after the core has reached the Schönberg–Chandrasekhar limit ($t_{\rm RLOF}/t_{\rm MS} > 1$). These systems never reach short periods, because mass loss causes the donor's radius to increase rather than decrease. The orbit widens in response to mass loss, leading to further expansion and mass loss. Such systems eventually form helium white dwarfs, with a well-characterized relation between final orbital period and mass \citep[e.g.][]{Rappaport1995}. The division between such ``divergent'' systems and converging systems that become CVs is called the ``bifurcation limit'' and has been studied elsewhere \citep[][]{Pylyser1988, Podsiadlowski_2003}.
    \item {\it Higher or lower initial donor masses:} For significantly lower initial donor masses, the donor cannot terminate its main-sequence evolution within a Hubble time, excluding the formation of a CV with an evolved donor. For higher initial masses, the donor has a radiative envelope on the main sequence, making magnetic braking inefficient. This does not qualitatively change the system's evolution once mass transfer begins, because mass loss leads to a drop in $T_{\rm eff}$ and efficient magnetic braking. However, due to inefficient magnetic braking and a shorter main-sequence lifetime, donors with higher initial masses are able to reach short periods and become CVs only if the initial period is short (e.g., $P_{\rm init}\lesssim 1$ day for an initial donor mass of $2 M_{\odot}$). In addition, the mass transfer rates shortly after RLOF are much higher (up to $10^{-6} M_{\odot}\,\rm yr^{-1}$) in models with more massive donors. This allows for stable hydrogen burning on the surface of the massive WD, potentially producing a supersoft X ray source or SNe Ia \citep{vandenHeuvel1992, Han2004}.
    \item {\it Different magnetic braking law}: We also experimented with models in which magnetic braking is not turned off when the donor's envelope becomes radiative, as is currently the default behavior in MESA. The most significant difference between these models and those presented in Figures~\ref{fig:evol_mesa}-\ref{fig:p_vs_m} is that their orbital decay does not slow down when the donor heats up. As a result, even the most evolved models that reach short periods remain mass transferring (though at low rates; $\dot{M} < 10^{-10} M_{\odot}\,\rm yr^{-1}$).
    
\end{itemize}

\section{Summary and Discussion}
\label{sec:discussion}

We have presented discovery data for LAMOST J0140355+392651 (J0140), an unusual binary with period $P_{\rm orb} = 3.81 $ hours containing a white dwarf and a hot, low-mass secondary. Our main conclusions are as follows:

\begin{enumerate}
    \item {\it Basic properties}: The system's optical light curve is dominated by ellipsoidal variability (Figure~\ref{fig:ztf_lcs}). The optical spectrum and SED of J0140 are dominated by the low-mass star, which appears to be a proto-WD (Figures~\ref{fig:spectra} and~\ref{fig:SED}). The massive WD dominates in the UV. Fitting the SED, {\it Gaia} parallax, optical spectrum, and radial velocities allows us to constrain physical properties of the system, including the proto-WD's mass, radius, and temperature, the WD mass, the orbital inclination, and the distance (Figure~\ref{fig:corner} and Table~\ref{tab:constraints}). 
    
    \item {\it Comparison to CVs and ELM WDs}: The properties of J0140 are transitional between known populations of CVs and ELM WDs. The effective temperature of the secondary, $T_{\rm eff}=6800\pm 100$, is much higher than in any known CVs at similar periods: most CVs near $P_{\rm orb}=4$ hours have $T_{\rm eff} \sim 3500$\,K, and the hottest donors discovered previously had $T_{\rm eff} \sim 5000$\,K (Figure~\ref{fig:other_cvs}). The secondary's mass and radius are also smaller than those of normal CV donors at comparable period (Figure~\ref{fig:evol_mesa}). The system has not been observed in outburst; if it does experience outbursts, their recurrence time is likely at least a few years (Figures~\ref{fig:full_lc} and~\ref{fig:outburst_detec_prob}), implying that the mass transfer rate is very low. On the other hand, the secondary is cooler and more bloated than known ELM WD's (Figure~\ref{fig:hrd_mesa}).
    
    \item {\it Evolutionary history}: We calculated a set of MESA binary evolution models to investigate the formation and future evolution of J0140 (Figures~\ref{fig:evol_mesa}-\ref{fig:p_vs_m}). These models reproduce the secondary's high temperature, small mass and radius, and low mass transfer rate if the secondary overflowed its Roche lobe {\it just} at the end of its main-sequence lifetime (within a few percent). In this case, all that is left of the secondary at $P_{\rm orb} \sim 4$ hours is a helium core with a thin hydrogen-burning envelope. Some of these models remain as mass-transferring CVs for the rest of their evolution, while others become detached as they grow hotter and temporarily appear as detached ELM WDs  (Figure~\ref{fig:hrd_mesa}). Unlike normal CVs, which reach a minimum period of $\approx 80$ minutes, models with evolved donors reach much shorter minimum periods (5-30 minutes), where they will appear as AM CVn binaries (Figure~\ref{fig:p_vs_m}).
    
\end{enumerate}

Further observations are necessary to better understand the nature of the system. In particular, higher-resolution and higher-SNR spectra will allow a deeper search for emission features associated with accretion, which will enable a more conclusive determination of whether there is ongoing mass transfer.  UV observations with the {\it Hubble Space Telescope} will provide direct constraints on the temperature of the WD. Such observations will also constrain the nitrogen and carbon abundance of the secondary \citep[e.g.][]{Gainsicke2003, Ashley2020}. This will test the evolutionary scenario proposed in Section~\ref{sec:mesa}, which predicts that both the donor and WD should bear evidence of CNO processing: nitrogen and helium enhancement, and carbon depletion. Finally, higher-cadence photometry will characterize the short-timescale variability that is thus far manifest only as additional scatter in the light curve (Figure~\ref{fig:deltamag}). If this variability is due to pulsations, it will be useful for probing the internal structure of the WD \citep[e.g.][]{Hermes2012}.

J0140 is clearly a rare system, containing the hottest secondary among of order 1000 spectroscopically confirmed CVs with $P_{\rm orb} \lesssim 6$ hours. We expect, however, that there are numerous similar systems with comparable or brighter magnitude yet to be discovered. J0140 exhibits neither outbursts nor strong emission lines, the two observables through which CVs are most commonly identified. Existing samples of CVs are thus likely biased against objects like J0140. We suspect that similar objects can be efficiently identified by searching the light curves of sources that fall blueward of the main sequence in the color-magnitude diagram (e.g. Figure~\ref{fig:cmd}) for large-amplitude ellipsoidal variation. Light curves will be dominated by ellipsoidal variation only if the donor dominates the optical light, which at short periods will only occur if the donor is unusually hot. We have initiated such a search and will present its yields in future work.

\subsection{Connection to EL CVn binaries and sdA stars}

We note that the proto-WD in J0140 is in a similar evolutionary state to the bloated stripped stars found in EL CVn-type binaries \citep[e.g.][]{Maxted2014}. In EL CVn binaries, the companions are main-sequence stars rather than WDs. Like the ELM WDs discovered by the ELM survey, the stripped stars in known EL CVn binaries are somewhat hotter and more compact than the proto-WD in J0140, likely because they are further evolved. 

Another population to which J0140 is closely related is the ``sdA'' stars, a heterogeneous collection of stars discovered by \citet{Kepler2016} with A-type spectra that appeared to have higher surface gravities than main-sequence stars of similar temperature. A majority of the sdA stars turned out to be main–sequence stars with overestimated $\log g$  \citep[e.g.][]{Hermes2017}. However, a small fraction appear to be genuine pre-ELM WDs, and some of these have temperatures and surface gravities similar to the proto-WD in J0140 \citep{Pelisoli2019b}. This raises the question whether their evolutionary status is similar to that of J0140. 

To investigate this possibility, we analyzed the ZTF light curves of the 50 ``high probability'' pre-ELM WDs identified by \citet{Pelisoli2019b} within the SDSS sdA sample. 42 of these 50 sources have high-quality ZTF light curves. Only one of them shows clear high-amplitude variability,\footnote{SDSS J 132713.01+382514.0, with reported $T_{\rm eff} = 7967\pm 18$\,K, shows what appears to be ellipsoidal variability with $P_{\rm orb}=2.00$ hours and peak-to-peak variability amplitude $\approx 8$\% (compared to $\approx 25\%$ in J0140). The variability is also consistent with a reflection effect binary with $P_{\rm orb}=1.00$ hours.} and while likely somewhat tidally distorted, it is not close to filling its Roche lobe. We therefore conclude that the sdA pre-ELM WD candidates identified by \citet{Pelisoli2019b} have longer orbital periods than J0140, have already shrunk well within their Roche lobes, and likely formed from donors that initiated mass transfer {\it after} the end of their main-sequence evolution.

\subsection{Is J0140 detached or mass-transferring?}
\label{sec:disc_detached}
The lack of outbursts and strong emission lines in J0140 raises the question whether the system is a detached binary or a CV. Some of the MESA models we calculate (Figure~\ref{fig:evol_mesa}) do indeed become detached at $P_{\rm orb}\lesssim 4$ hours, because magnetic braking becomes inefficient after the donor loses its convective envelope. 

The relatively high temperature of the more massive WD inferred from the FUV emission ($T_{\rm eff}\approx 20,000$\,K) suggests that if the binary it detached, it must have ceased mass transfer recently.\footnote{This $T_{\rm eff}$ is really an upper limit, since some of the FUV emission could come from a disk. But a disk could not be invoked if the binary were detached.} Similarly, the large observed amplitude of ellipsoidal variation implies that the proto-WD is very close to Roche lobe overflow: our light curve fit (Section~\ref{sec:lc}) yields a Roche lobe filling factor $R_{\rm proto\,WD}/R_{L} > 0.98$. It is possible that detachment occurred recently and the donor is still almost Roche lobe filling, but this would require some fine tuning.

Long outburst recurrence timescales are expected at low accretion rate, because a longer period of sustained accretion is required to build up enough mass in the disk for it to become unstable \citep[e.g.][]{Warner_2003, Hameury_2020}. 
The intrinsic distribution of CV outburst timescales is difficult to constrain observationally, because there is a strong detection bias toward systems with frequent outbursts \citep[e.g.][]{Breedt_2014}.  Recently, \citet{Yu_2019} discovered a long-period CV ($P_{\rm orb}\sim 9$ hours) with only one outburst in 4 years of nearly continuous Kepler data. The expected future evolution of that system is similar to what we infer for J0140. V1460 Her, another CV proposed to contain an evolved donor, experienced only two confirmed outbursts in 14 years \citep[][]{Ashley2020}. Since J0140 appears to be the most extreme member of the evolved CV population discovered so far, we cannot rule out that outbursts occur on a long recurrence timescale. 

Given the unusual nature of the system, it is also worth considering whether accretion could be occurring without {\it any} outbursts. At sufficiently low accretion rates, cooling is inefficient and a thin disk is not expected to form. For  J0140, the minimum accretion rate below which even the outer disk is in the radiatively inefficient regime is $\dot{M} \lesssim {\rm few} \times 10^{-11} M_{\odot}\,\rm yr$ \citep[e.g.][]{Menou1999}, which is plausible for the accretion rates predicted by the MESA models in Section~\ref{sec:mesa_mdoels}.

We conclude that J0140 is either currently undergoing mass transfer at an unusually low rate, or recently became detached, but further observations are needed to confidently distinguish between these possibilities. 

\section*{Acknowledgements}
We thank Boris Gänsicke and Tom Marsh for helpful comments. We are grateful to the staff at Lick Observatory for their assistance in obtaining the Kast spectra, and to the developers of \texttt{pypeit} and \texttt{Phoebe} for maintaining and making public the codes.
KE acknowledges support from an NSF graduate research fellowship and a Hellman fellowship from UC Berkeley. This research benefited from meetings supported by the Gordon and Betty Moore Foundation through grant GBMF5076. EQ was supported in part by a Simons Investigator award from the Simons Foundation. KJS received support for this work from NASA through the Astrophysics Theory Program (NNX17AG28G). We thank Geoff Tabin and In-Hei Hahn for their hospitality during the writing of this manuscript. 

%%%%%%%%%%%%%%%%%%%%%%%%%%%%%%%%%%%%%%%%%%%%%%%%%%
\section*{Data Availability}

Data in this paper are available upon reasonable request to the corresponding author.



\bibliographystyle{mnras}

%%%%%%%%%%%%%%%%%%%%%%%%%%%%%%%%%%%%%%%%%%%%%%%%%%

%%%%%%%%%%%%%%%%% APPENDICES %%%%%%%%%%%%%%%%%%%%%

\appendix

\section{Summary of LAMOST data}
\label{sec:spec_details}

Table~\ref{tab:lamost} lists the parameters of J0140 as reported in LAMOST DR5 and measured from combined spectra. These spectra are expected to be subject to significant orbital smearing, so parameters derived from them should be interpreted with caution. 

\begin{table}
	\centering
	\caption{LAMOST DR5 parameters for J0140, estimated from spectra obtained in two separate visits. A third visit (\texttt{obsid} 1012095) had too-low SNR to yield useful RVs or stellar parameters. }
	\label{tab:lamost}
	\begin{tabular}{cccc} 
		\hline
		parameter & units & visit 1 & visit 2 \\
		\hline
		obsid & -- & 184214089 & 353912095 \\
		mjd & d & 56621 & 57278 \\
		$g$-band SNR & -- & 14.9 & 19.7 \\
		RV & $\rm km\,s^{-1}$ & $-5\pm 51$ & $-377\pm 62$ \\
		$T_{\rm eff}$ & K & $6586\pm246$ & $6394\pm 829$\\
		$\log (g/(\rm cm\,s^{-2}))$ & -- & $4.37\pm0.39$ & $4.22\pm 1.19$\\
		$\rm [Fe/H]$ & dex & $-0.17\pm 0.23$ & $-0.16\pm 0.77$\\
        spectral\,\,type & -- & F0 & F3 \\

		\hline
	\end{tabular}
\end{table}

\begin{figure}
    \centering
    \includegraphics[width=\columnwidth]{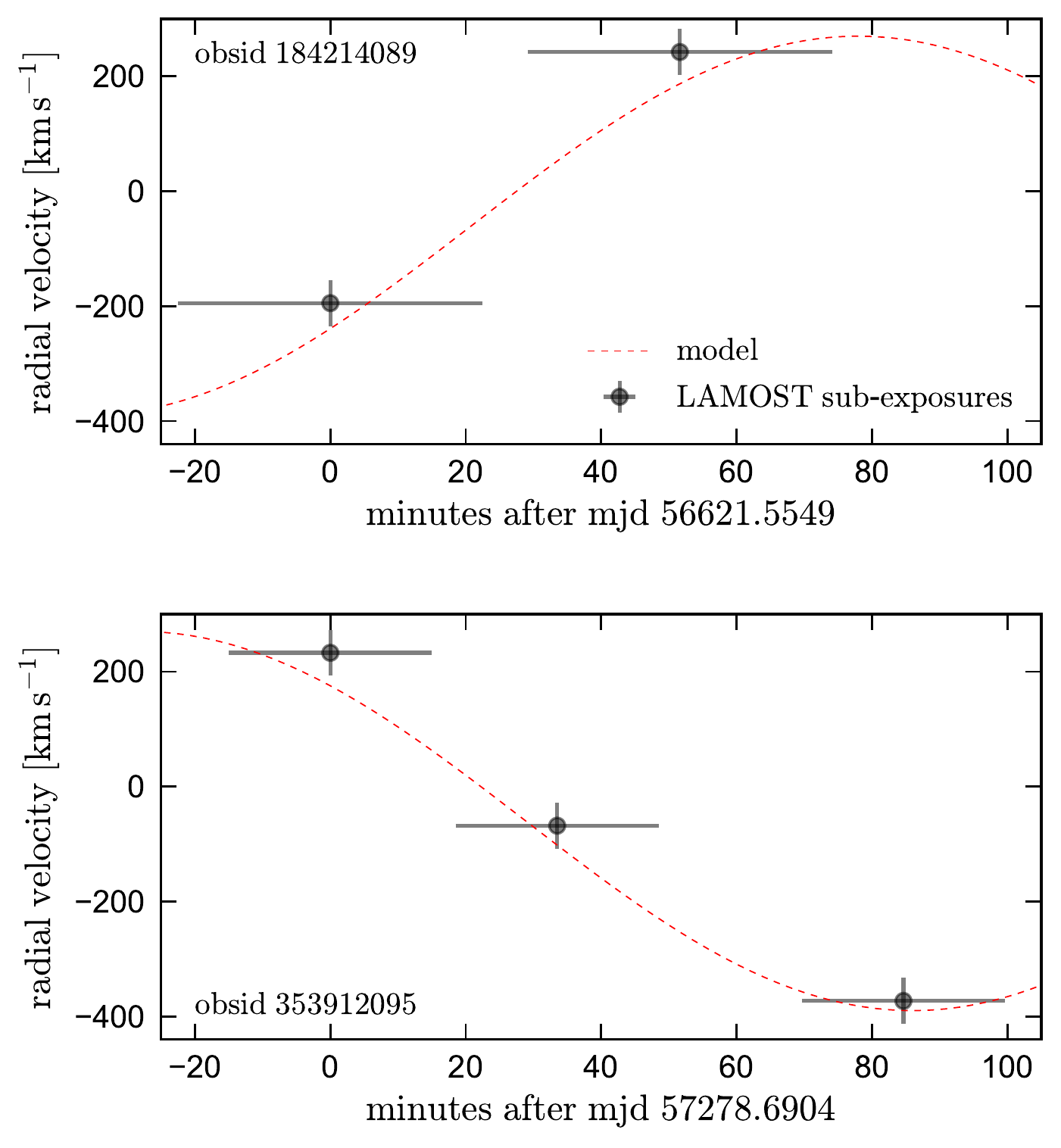}
    \caption{Radial velocities measured from individual LAMOST sub-exposures for two separate visits. Horizontal errorbars show exposure time. Sub-exposure spectra are normally coadded (without any RV shifts) to produce the visit spectra from which stellar parameters are derived. For J0140, there are large RV shifts between visits, so the combined spectrum is subject to orbital smearing. Red line shows the radial velocity predicted by the orbital solution obtained from the light curve.}
    \label{fig:lamost_rvs}
\end{figure}

The first visit (\texttt{obsid} 184214089) was split between two 2700 second subexposures, while the second (\texttt{obsid} 353912095)  was split between three 1800 second subexposures. RVs measured from the subexposures are shown in Figure~\ref{fig:lamost_rvs}. There is clearly significant RV variation (up to 600\,$\rm km\,s^{-1}$) between subexposures of a single visit, implying that the spectra produced by coadding these sub-exposures will be smeared out. Indeed, the sub-exposure spectra are also subject to significant smearing, as the RV varies by more than $300\,\rm km\,s^{-1}$ between adjacent sub-exposures. 

Figure~\ref{fig:lamost_rvs} also shows the RVs predicted for the orbital ephemeris derived in Section~\ref{sec:ephemeris} from the light curve and our follow-up spectroscopy. The LAMOST RVs are not used in deriving the ephemeris, but they are in reasonably good agreement with it. All 5 subexposure RVs are offset above our orbital solution, by $\approx 40\,\rm km\,s^{-1}$ on average. This may indicate a zeropoint offset in our orbital solution compared to the LAMOST RV scale, since our follow-up RVs are not absolutely calibrated. This would have no effect on the inferred physical parameters of the binary.  

\section{Summary of Kast spectra}
\label{sec:kast_summary}
Table~\ref{tab:specs} summarizes the multi-epoch spectra of J0140 we obtained with the Kast spectrograph at Lick observatory.

\begin{table}
	\centering
	\caption{Summary of Kast spectra. HMJD is mid-exposure time. Setup is described in Section~\ref{sec:kast}. }
	\label{tab:specs}
	\begin{tabular}{cccccc} 
		\hline
		HMJD UTC & phase & exptime [s] & setup & SNR & RV [$\rm km\,s^{-1}$] \\
		\hline
59060.3371 & 0.04 & 600 & B & 5.6 & $-118\pm 21$  \\
59060.3444 & 0.08 & 600 & B & 7.9 & $-176\pm 19$  \\
59060.3518 & 0.13 & 600 & B & 8.8 & $-296\pm 13$  \\
59060.3643 & 0.21 & 600 & B & 10.3 & $-378\pm 12$  \\
59060.3717 & 0.25 & 600 & B & 11.7 & $-378\pm 10$  \\
59060.3790 & 0.30 & 600 & B & 11.3 & $-386\pm 12$  \\
59060.3940 & 0.39 & 600 & B & 10.8 & $-243\pm 12$  \\
59060.4015 & 0.44 & 600 & B & 10.1 & $-137\pm 12$  \\
59061.4040 & 0.76 & 600 & A & 11.5 & $311\pm 19$  \\
59061.4113 & 0.81 & 600 & A & 10.4 & $232\pm 30$  \\
59061.4186 & 0.85 & 600 & A & 10.9 & $246\pm 20$  \\
59061.4305 & 0.93 & 600 & A & 9.3 & $105\pm 21$  \\
59061.4378 & 0.97 & 600 & A & 8.8 & $32\pm 24$  \\
59062.3340 & 0.62 & 600 & A & 10.4 & $188\pm 16$  \\
59062.3414 & 0.67 & 600 & A & 10.5 & $274\pm 33$  \\
59062.3487 & 0.72 & 600 & A & 9.3 & $260\pm 19$  \\
59062.3601 & 0.79 & 600 & A & 10.5 & $239\pm 42$  \\
59062.3674 & 0.83 & 600 & A & 9.6 & $189\pm 55$  \\
59062.3747 & 0.88 & 600 & A & 8.5 & $-15\pm 54$  \\
59064.3512 & 0.34 & 600 & A & 5.4 & $-287\pm 56$  \\
59064.3585 & 0.39 & 600 & A & 6.0 & $-301\pm 54$  \\
59064.3658 & 0.43 & 600 & A & 6.4 & $-132\pm 38$  \\
59064.4701 & 0.09 & 600 & A & 13.6 & $-215\pm 32$  \\
59064.4774 & 0.14 & 600 & A & 13.9 & $-276\pm 24$  \\
59069.3438 & 0.81 & 600 & A & 17.6 & $240\pm 12$  \\
59069.3511 & 0.86 & 600 & A & 16.9 & $194\pm 13$  \\
59069.3584 & 0.90 & 600 & A & 15.2 & $70\pm 13$  \\
59069.3708 & 0.98 & 600 & A & 10.5 & $-32\pm 21$  \\
59069.3781 & 0.03 & 600 & A & 13.5 & $-155\pm 16$  \\
59069.3854 & 0.07 & 600 & A & 13.6 & $-232\pm 12$  \\
59069.3927 & 0.12 & 600 & A & 14.5 & $-315\pm 18$  \\
59069.4047 & 0.20 & 600 & A & 16.0 & $-355\pm 15$  \\
59069.4120 & 0.24 & 600 & A & 16.3 & $-391\pm 15$  \\
59069.4193 & 0.29 & 600 & A & 16.4 & $-361\pm 13$  \\
59069.4266 & 0.33 & 600 & A & 15.8 & $-348\pm 20$  \\
59069.4385 & 0.41 & 600 & A & 14.9 & $-208\pm 19$  \\
59069.4458 & 0.45 & 600 & A & 14.7 & $-146\pm 7$  \\
59069.4531 & 0.50 & 600 & A & 14.6 & $-58\pm 10$  \\
59069.4604 & 0.55 & 600 & A & 14.3 & $57\pm 10$  \\
59069.4715 & 0.62 & 600 & A & 14.8 & $179\pm 10$  \\
59069.4788 & 0.66 & 600 & A & 14.3 & $257\pm 20$  \\
		\hline
	\end{tabular}
\end{table}

\section{Phase-dependent spectra}
\label{sec:phased_specs}
\begin{figure}
    \centering
    \includegraphics[width=\columnwidth]{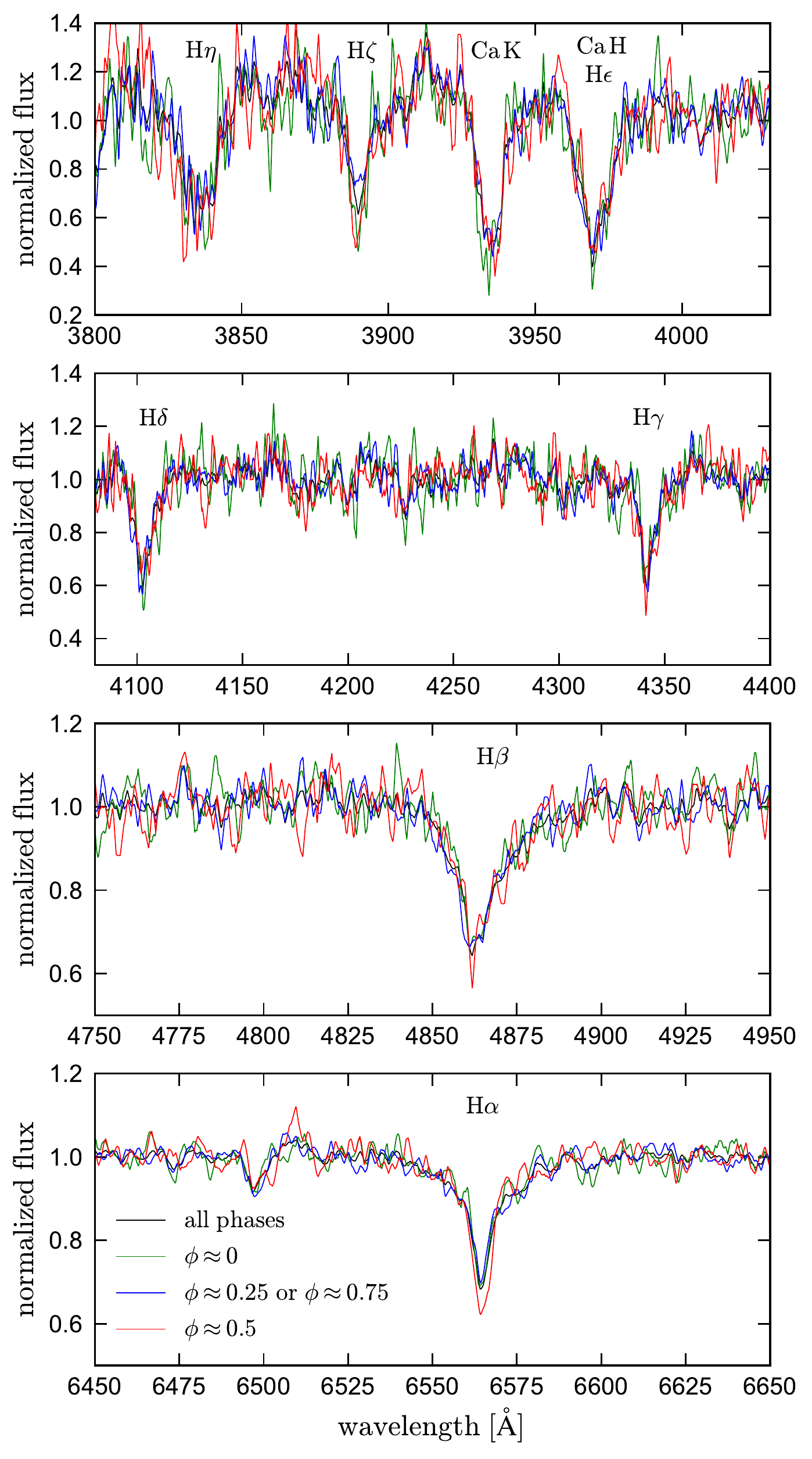}
    \caption{Coadded spectra of J0140, produced by stacking single-epoch spectra with phases within 0.1 of the values specified in the legend. The strongest lines are labeled. No strong phase dependence is obvious, but there are hints of increased H$\alpha$ absorption near $\phi=0.5$. This may be due to the full or partial eclipse of a H$\alpha$ emitting component at $\phi=0.5$.}
    \label{fig:line_var}
\end{figure}

To investigate the possibility of variability in the spectra with orbital phase, we binned the single-epoch spectra in phase before coadding. Figure~\ref{fig:line_var} compares the coadded spectra at $\phi \approx 0$ (when the WD passes in front of the donor), $\phi \approx 0.5$ (when the donor passes in front of the WD), and  $\phi \approx 0.25$ or  $\phi \approx 0.75$ (when both components are viewed side-on). In all cases, we include spectra within 0.1 of the indicated phase. Most lines have consistent strengths  across phase bins. The H$\alpha$ line, however, appears somewhat stronger at $\phi \approx 0.5$ than at other phases. This is the phase at which the WD and disk (if it exists) are behind the donor, so the increased H$\alpha$ absorption at $\phi \approx 0.5$ is consistent with a scenario in which the disk contributes weak H$\alpha$ emission that is fully or partially eclipsed at $\phi \approx 0.5$. An alternative explanation is that irradiation by the hot WD gives rise to emission on one side of the donor, which would be hidden from view at $\phi \approx 0.5$.  Higher-SNR spectra are required to confirm the phase-dependence of the H$\alpha$ line, and, if it is confirmed, to further elucidate its origin.

\section{CMD position}
\begin{figure*}
    \centering
    \includegraphics[width=\textwidth]{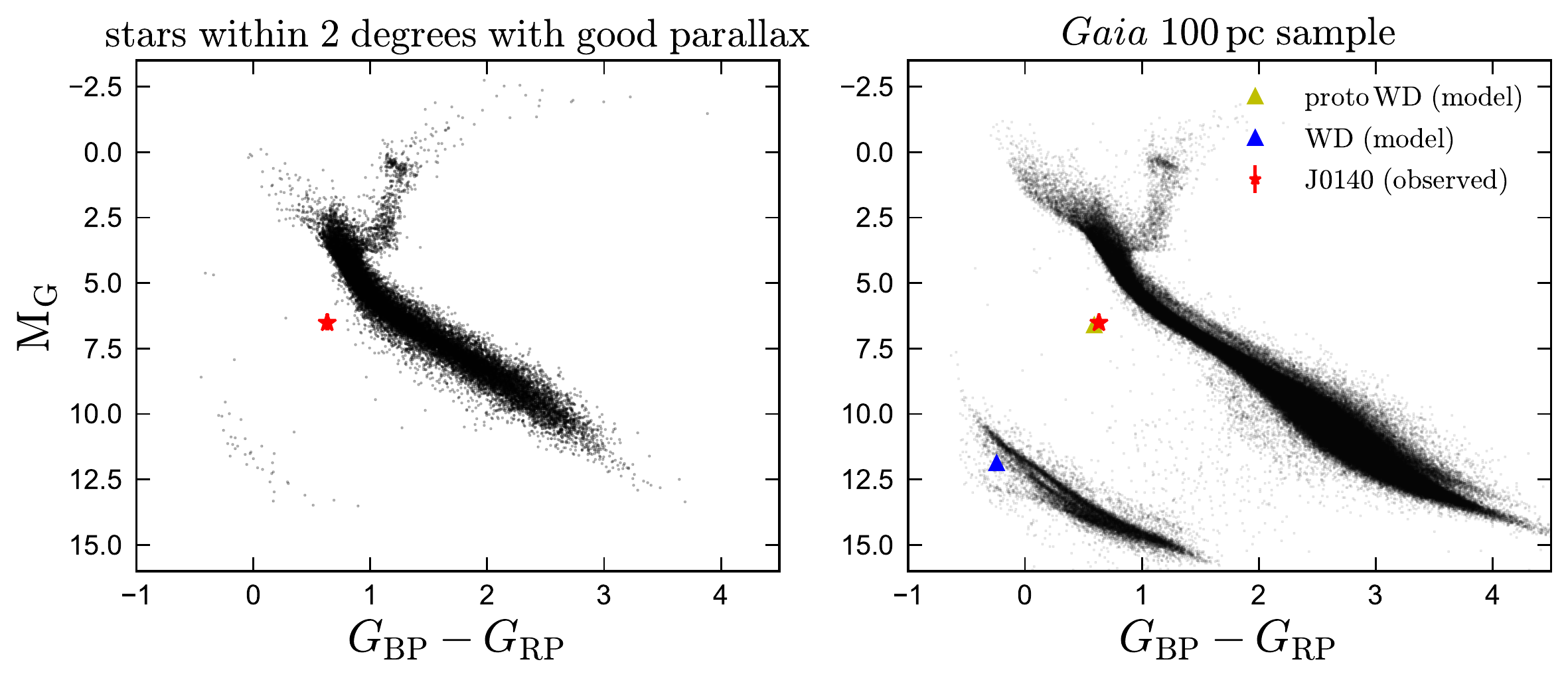}
    \caption{J0140 on the {\it Gaia} color-magnitude diagram (CMD). In the left panel, black points show sources within 2 degrees of J0140 with clean and precise photometry and astrometry; in the right panel, they show sources over the whole sky but with $d < 100$\,pc. J0140 is significantly bluer than normal main-sequence stars of the same absolute magnitude. Right panel shows the predicted individual CMD positions of the proto-WD and WD from the SED models in Figure~\ref{fig:SED}, including effects of extinction. Even without the light contributions of the WD, the donor is off the main sequence. }  
    \label{fig:cmd}
\end{figure*}

Figure~\ref{fig:cmd} shows J0140 on the {\it Gaia} eDR3 color-magnitude diagram. We compare it to other stars with good astrometry within 2 degrees (which are subject to a similar extinction) and to the {\it Gaia} 100 pc sample.\footnote{For the 100 pc sample, we adopt the photometric and astrometric cuts suggested by \citet{Lindegren_2018}. For the stars within 2 degrees, we use the same quality cuts but only require \texttt{parallax\_over\_error} > 5 and \texttt{parallax} > 0.5.} J0140 falls below the main sequence; that is, it is bluer than a main-sequence star of the same absolute magnitude. In the right panel, we show the predicted individual CMD positions of the donor, disk, and WD. As in Figure~\ref{fig:SED}, we caution that the SED of the disk and WD are poorly constrained.

Curiously, the CMD position of J0140 is quite similar to that of ``normal'' CVs with the same period \citep[e.g.][]{Abril2020, Abrahams2020}. Most CVs with $P_{\rm orb}\sim 4$ hours, however, are completely dominated by the accretion disk in the optical, even in quiescence. At $P_{\rm orb}\sim 4$ hours, the total luminosity of donors on the CV donor sequence is only  $0.015 L_{\odot}$ \citep{Knigge_2006}, about $10\%$ of the luminosity of the donor in J0140. In J0140, the donor dominates. The system has a lower accretion rate -- and correspondingly, a fainter disk -- than ordinary CVs of its period. This compensates for its unusually high donor luminosity, leading to an unremarkable  total absolute magnitude.

%%%%%%%%%%%%%%%%%%%%%%%%%%%%%%%%%%%%%%%%%%%%%%%%%%

% Don't change these lines
\bsp	% typesetting comment
\label{lastpage}
\end{document}